%% file: approxis.tex
\pdfoutput=1
\RequirePackage{etoolbox}
\newbool{fullversion}
\booltrue{fullversion}
\newbool{arxivversion}
\booltrue{arxivversion}

\RequirePackage{etoolbox}

\providebool{fullversion}
\providebool{arxivversion}

\newbool{needspace}
\boolfalse{needspace}

\ifbool{false}{%
\documentclass[acmsmall,screen,nonacm]{acmart}\settopmatter{printfolios=true,printccs=false,printacmref=false}
}{%
  \documentclass[acmsmall,screen]{acmart}\settopmatter{}
}
\usepackage{preamble}
\usepackage{iris}
\usepackage{prob}
\usepackage{examples}

\setcopyright{rightsretained}
\acmDOI{10.1145/3704877}
\acmYear{2025}
\acmJournal{PACMPL}
\acmVolume{9}
\acmNumber{POPL}
\acmArticle{41}
\acmMonth{1}
\received{2024-07-11}
\received[accepted]{2024-11-07}

\ifbool{fullversion}{
  \newcommand{\appref}[1]{\cref{#1}}
  \newcommand{\Appref}[1]{\Cref{#1}}
}{
  \newcommand{\appref}[1]{Appendix A of \cite{approxis:arxiv}}
  \newcommand{\Appref}[1]{Appendix A of \cite{approxis:arxiv}}
}

\title{Approximate Relational Reasoning for Higher-Order Probabilistic Programs}

\author[P. G. Haselwarter]{Philipp~G. Haselwarter}
\orcid{0000-0003-0198-7751}
\affiliation{
  \institution{Aarhus University}
  \country{Denmark}
}
\email{pgh@cs.au.dk}

\author[K. H. Li] {Kwing Hei Li}
\orcid{0000-0002-4124-5720}
\affiliation{
  \institution{Aarhus University}
  \country{Denmark}
}
\email{hei.li@cs.au.dk}

\author[A. Aguirre]{Alejandro Aguirre}
\orcid{0000-0001-6746-2734}
\affiliation{
  \institution{Aarhus University}
  \country{Denmark}
}
\email{alejandro@cs.au.dk}

\author[S. O. Gregersen]{Simon Oddershede Gregersen}
\orcid{0000-0001-6045-5232}
\affiliation{
  \institution{New York University}
  \country{USA}
}
\email{s.gregersen@nyu.edu}

\author[J. Tassarotti]{Joseph Tassarotti}
\orcid{0000-0001-5692-3347}
\affiliation{
  \institution{New York University}
  \country{USA}
}
\email{jt4767@cs.nyu.edu}

\author[L. Birkedal]{Lars Birkedal}
\orcid{0000-0003-1320-0098}
\affiliation{
  \institution{Aarhus University}
  \country{Denmark}
}
\email{birkedal@cs.au.dk}

\ifbool{false}{
\usepackage{caption}
}{
\usepackage[belowskip=-10pt,aboveskip=4pt]{caption}
}

\makeatletter
\ifbool{needspace}{
\let \MathparLineskip \mpr@lesslineskip %
}{}
\makeatother

\ifbool{needspace}{
  \newcommand{\vsquish}[1]{\vspace{-#1}}
}{
  \newcommand{\vsquish}[1]{}
}

\begin{document}

\begin{abstract}
  Properties such as provable security and correctness for randomized programs are naturally expressed relationally as approximate equivalences.
  As a result, a number of relational program logics have been developed to reason about such approximate equivalences of probabilistic programs.
  However, existing approximate relational logics are mostly restricted to first-order programs without general state.

  In this paper we develop \theaplog, a \emph{higher-order approximate relational separation logic} for reasoning about approximate equivalence of programs written in an expressive ML-like language with discrete probabilistic sampling, higher-order functions, and higher-order state.
  The \theaplog~logic recasts the concept of \emph{error credits} in the relational setting to reason about relational approximation, which allows for expressive notions of modularity and composition, a range of new approximate relational rules, and an internalization of a standard limiting argument for showing exact probabilistic equivalences by approximation. We also use \theaplog to develop a logical relation model that quantifies over error credits, which can be used to prove \emph{exact contextual equivalence}.
  We demonstrate the flexibility of our approach on a range of examples, including the PRP/PRF switching lemma, IND\$-CPA security of an encryption scheme, and a collection of rejection samplers.
  All of the results have been mechanized in the Coq proof assistant and the Iris separation logic framework.
\end{abstract}



\maketitle

\input{introduction.tex}

\input{key-ideas.tex}

\input{preliminaries.tex}

\input{logic.tex}

\input{logrel.tex}

\input{case-studies.tex}

\input{model.tex}

\input{work.tex}

\input{conclusion.tex}

\begin{acks}
  This work was supported in part by the \grantsponsor{NSF}{National Science Foundation}{}, grant no.~\grantnum{NSF}{2338317}, the \grantsponsor{Carlsberg Foundation}{Carlsberg Foundation}{}, grant no.~\grantnum{Carlsberg Foundation}{CF23-0791}, a \grantsponsor{Villum}{Villum}{} Investigator grant, no. \grantnum{Villum}{25804}, Center for Basic Research in Program Verification (CPV), from the VILLUM Foundation, and the European Union (\grantsponsor{ERC}{ERC}{}, CHORDS, \grantnum{ERC}{101096090}).
  Views and opinions expressed are however those of the author(s) only and do not necessarily reflect those of the European Union or the European Research Council.
  Neither the European Union nor the granting authority can be held responsible for them.
\end{acks}

\ifbool{fullversion}{
\appendix
\input{appendix-case-studies.tex}

\FloatBarrier
}{}
\bibliography{refs}
\vfill

\end{document}

%% file: introduction.tex
\section{Introduction}
\label{sec:introduction}

Many important properties of probabilistic programs are naturally expressed as \emph{approximate} equivalence of two programs.
For example, provable security~\cite{DBLP:journals/jcss/GoldwasserM84} compares an implementation of a cryptographic scheme to an idealized specification program that does not have access to any sensitive information, and aims to show that an adversary can only distinguish them with some small probability.
In a similar spirit, many randomized algorithms and data structures can be specified by showing that they are approximately equivalent to their non-probabilistic counterparts.
Consequently, it is important to be able to reason about approximate equivalences and so a number of relational program logics have been developed for first-order languages \cite{apRHL, apRHL+, EpRHL} or higher-order languages with first-order global state \cite{HORHL}.

In this work, we develop \theaplog, a \emph{higher-order approximate relational separation logic} for reasoning about approximate equivalence of \thelang programs, an expressive ML-like language with discrete random sampling, higher-order functions, and higher-order dynamically-allocated state.
A key point is that \theaplog, inspired by the unary Eris logic \cite{eris}, introduces \emph{error credits} in the relational setting to reason about approximation.
Error credits are separation-logic resources that bound the maximum approximation error between two programs.
We introduce a collection of novel \emph{approximate coupling rules}, which consume error credits in order to relate randomized transitions of two programs.
By treating the relational approximation error as just another separation-logic resource, \theaplog provides modular reasoning principles that enable more precise error accounting when composing proofs, much as Eris demonstrated in the non-relational setting.

Surprisingly, error credits not only allow us to prove approximate equivalences, they also allow us to prove \emph{exact} equivalences that were beyond the scope of prior coupling-based relational program logics.
Just as in real analysis, where one can prove two numbers are equal by showing that the distance between them is smaller than $\err$ for all $\err > 0$, we can similarly show two probability distributions are equivalent by showing the distance between them is bounded by $\err$ for all $\err > 0$.
Using \theaplog, we show how to recover this technique internally in the logic through \emph{error amplification}~\cite{eris} and thus prove exact equivalence of probabilistic programs by means of approximation.
Based on this, we develop a new binary logical relations model of a rich type system for \thelang{} with recursive types and impredicative polymorphism.
The model supports approximate reasoning and gives us a powerful and novel method for showing exact \emph{contextual} equivalence of higher-order probabilistic programs. For other existing approaches, including both operational approaches, \eg{}, Clutch \cite{clutch}, and denotational approaches, \eg{}, pRHL \cite{pRHL} and HO-RHL \cite{HORHL}, some of the examples that we consider would be very complicated---if not impossible---to handle.

We show that \theaplog scales to more involved approximate reasoning by showing the classical PRP/PRF Switching Lemma \cite{PRPPRF, PRPPRF2} and IND\$-CPA security of a PRF-based symmetric encryption scheme.
Moreover, we apply error amplification and our logical relation to show contextual equivalences for a collection of rejection samplers, including a sampling scheme for drawing a random sample from a B+ tree~\citep{b+_tree}.

Examples like the PRP/PRF Switching Lemma have been verified in many different settings, but we emphasize the rich programming language we consider here.
While some of these examples might be expressible in simpler languages, features such as higher-order functions, higher-order state, and polymorphism are all found in general-purpose programming languages, and are needed for modern compositional software development.
Moreover, cryptographic security can be more naturally expressed in such higher-order languages and avoids the need for syntactic restrictions on adversaries as seen, \eg{}, in EasyCrypt \cite{easycrypt}.
As a consequence, verification frameworks must handle these language features to reason about large applications and realistic implementations.
Higher-order separation logic is a powerful and well-tested abstraction for this purpose, and \theaplog{} shows how to beneficially apply it for approximate relational reasoning.
While the B+ tree case study, for example, is quite involved, the complexity is managed through mostly-standard separation-logic reasoning.
We see this as a significant strength of our approach.

At a technical level, our development builds upon the (non-approximate) probabilistic coupling logic Clutch \cite{clutch}.
By incorporating error credits \cite{eris} in the relational setting, our development generalizes the approach to approximate reasoning using approximate couplings.
In addition, we introduce two new \emph{coupling precondition} connectives and a notion of \emph{erasability}.
The erasability condition not only captures the soundness of asynchronous couplings \cite{clutch} in a more semantic way, but also allows for a more principled approach to validating the new approximate and non-approximate coupling rules we introduce and which are critical for the examples that we consider.

\paragraph{Contributions} In summary, we make the following contributions:
\begin{itemize}
\item The first higher-order approximate relational separation logic, \theaplog, for reasoning about approximate equivalence of \thelang programs, an expressive ML-like language with probabilistic sampling, higher-order functions, and higher-order state,
\item A logical internalization of a limiting argument that allows us to show exact equivalence of higher-order probabilistic programs through approximation,
\item A class of new approximate and non-approximate coupling rules, including the \emph{many-to-one} and \emph{fragmented} coupling rules,
\item A logical relations model of an expressive type system for \thelang{} with recursive types and impredicative polymorphism, which allows us to show (exact) \emph{contextual} equivalence of probabilistic programs through a limiting argument,
\item A collection of case studies: the PRP/PRF Switching Lemma \cite{PRPPRF, PRPPRF2}, IND\$-CPA security of a PRF-based symmetric encryption scheme, and contextual equivalence of a selection of rejection samplers, including a sampling scheme for drawing a random sample from a B+ tree~\citep{b+_tree}.
  Several of these are, to the best of our knowledge, beyond the scope of previous techniques, in particular for expressive languages such as \thelang{}. 
\item Full mechanization of all results in the Coq proof assistant \cite{coq}, building on top of the Iris separation logic framework~\cite{irisjournal} and the Coquelicot~\cite{coquelicot} library for real analysis. 
\end{itemize}

\paragraph{Outline}
In \cref{sec:key-ideas} we give high-level intuition for how to reason using \theaplog.
Here we discuss the PRP/PRF Switching Lemma, a classical result in cryptography, and show how to use the limiting argument on a simple rejection sampler.
In \cref{sec:preliminaries} we recall some definitions from probability theory and define the semantics of \thelang.
In \cref{sec:arl} present a collection of program logic rules and coupling rules of \theaplog before developing our logical relations model in \cref{sec:logrel}.
In \cref{sec:case-studies} we showcase \theaplog on a range of case studies, and in \cref{sec:model} we explain how the semantic model of \theaplog is constructed on top of the Iris base logic.
Finally, we discuss related work and conclude in \cref{sec:work} and \cref{sec:conclusion}, respectively.

%% file: key-ideas.tex
\section{Key Ideas}
\label{sec:key-ideas}

In this section, we give a high-level overview of \theaplog and introduce how error credits can be used to do approximate relational reasoning.
The primary specification assertion in \theaplog is the \emph{refinement weakest precondition}, written $\wpref{\expr_1}{\expr_2}{\Phi}$, where $\expr_1$ and $\expr_2$ are two randomized programs, and $\Phi$ is a relation on the return values and final program states of \(\expr_1\) and \(\expr_2\).
Informally, this relational connective says that if executing $\expr_1$ terminates with a value $\val_{1}$, then $\expr_2$ terminates with value $\val_{2}$ and the postcondition $\Phi(\val_{1}, \val_{2})$ holds.

Because \theaplog is a separation logic, when presenting the rules of the logic, we use inference rule-style notation with premises $\prop_{1}, \ldots, \prop_{n}$ and conclusion $\propB$ to stand for the entailment $\prop_{1} \sep \ldots \sep \prop_{n} \vdash \propB$ in the logic.

For reasoning about non-randomized steps of $e_1$ and $e_2$, \theaplog has a variety of rules that are relational generalizations of usual separation logic rules, as in prior relational Hoare logics~\citep{DBLP:conf/popl/Benton04, reloc, caresl}.
For randomized steps, the first tool \theaplog provides are the so-called \emph{coupling} rules pioneered by pRHL~\citep{pRHL}.
A simple, specialized form of such a rule is
\begin{equation*}
  \infrule[Right]{wp-couple-exact}
  { \All n \leq \tapebound . \wpref n n \Phi }
  { \wpref{\Rand \tapebound}{\Rand \tapebound}{\Phi} }
\end{equation*}
where $\Rand \tapebound$ is a command in the language that samples a value uniformly from $\{0, \dots, \tapebound\}$.
This rule says that if both programs are sampling from $\Rand \tapebound$, then we may reason \emph{as if} they both returned the same sample value $n$, instead of having to consider all $(N+1)^2$ possible combinations of values they could have returned.
This rule is justified by using the notion of couplings from probability theory, and relies on the fact that the two sets being sampled from have the same size.

What if we want to reason about the case where the two sets being sampled from are \emph{not} the same size?
For example, suppose the left program executes $\Rand N$ and the right executes $\Rand~(N + 1)$. 
We cannot exactly reason as if both programs sample the same value: there is a chance that the program on the right samples $N + 1$, which the program on the left can never do!
However, the right program only draws this ``bad'' value of $N+1$ with probability $1/(N+2)$.
If $N$ is very large, this probability will be small, so we might hope to argue that we can approximately reason as if the two samples returned the same value, recovering an analogue of \ruleref{wp-couple-exact}.

This idea of approximate relational reasoning has been developed in apRHL~\citep{apRHL}.
In apRHL, relational Hoare triples are annotated with an additional parameter, $\err$, which bounds the approximation error.\footnote{apRHL has a second annotation for bounding another form of probabilistic approximation which we do not consider.}
Then, the coupling rules allow for relating two sampling commands from distributions that are only equal up to some error $\err'$ by adding $\err'$ to the total error on the Hoare triple.
However, \citet{eris} have previously shown that tracking an error bound as an additional parameter of a Hoare judgement has a number of limitations related to modularity and precision of bounds.
Instead, they proposed to track errors through a separation logic assertion called an \emph{error credit}, written $\upto{\err}$, which represents a ``permission'' to incur an approximation error of up to $\err$. 
They developed this idea in a \emph{unary} logic called Eris for bounding the probabilities of events of a randomized program.
A key aspect of the flexibility of error credits arises from the fact that they can be split and joined, in the sense that $\upto{\err_1 + \err_2} \dashv\vdash \upto{\err_1} * \upto{\err_2}$ for $\err_1, \err_2 \geq 0$. 

\theaplog uses this idea of error credits to track approximation error in couplings.
A special case of \theaplog's approximate coupling rule applied to the scenario described above would be:
\begin{equation*}
  \infrule*[lab]{}
  {
    \upto{\tfrac{1}{\tapebound+2}}\\
    \All n \leq \tapebound .
    \wpref{n}{n}{\Phi}
  }
  {\wpref{\Rand \tapebound}{\Rand~(\tapebound + 1)}{\Phi}}
\end{equation*}
which says that if we spend $\upto{\tfrac{1}{\tapebound+2}}$ credits we may reason as if the two samples returned the same value.
Informally, we think of the error credits as being spent to ``rule out'' the case where the program on the right returns $N + 1$.

In \theaplog a derivation of the form $\upto{\err} \vdash \wpref{e_1}{e_2}{\Phi}$ implies that at most $\err$ total error is incurred in deriving the refinement weakest precondition.
The soundness theorem for the logic then says that to prove that the distributions corresponding to two programs are within $\err$ distance of one another (in a sense to be made precise later), it suffices to prove the refinement weakest preconditions in both directions, each with error up to $\err$. 

At a high level, tracking approximate coupling error using credits seems like a relatively simple adaptation of Eris~\cite{eris} to the relational setting.
However, as we shall see later, doing so in a sound manner involves addressing several new technical challenges that have no analogue in the unary case.
But first we shall look at an example of how the features of \theaplog can be used to reason about cryptographic security.

\subsection{Motivating Example: PRP/PRF Switching Lemma}
\label{sec:wsl}

To illustrate the different ideas coming together in \theaplog, we explore a classic approximate equivalence result from cryptography: the PRP/PRF Switching Lemma~\cite{PRPPRF,PRPPRF2}.
A key part of this lemma involves showing that random permutations (RPs) are hard to distinguish from random functions (RFs) by a client (the ``adversary'') that can only make a bounded number of queries to such functions.
For finite sets $X$ and $Y$, a random function $f : X \to Y$ can be sampled by selecting, for each $x \in X$, an independent, uniform sample from $Y$, to use as the value for $f(x)$.
Sometimes it is desirable for \(f\) to be invertible (for modeling encryption and decryption of a block cipher). We call such an \(f\) a random permutation.
The difference between a RP and a RF is then that a RF may have collisions, \ie, values \(x_1 \not= x_2\) such that \(f(x_1) = f(x_2)\), while a RP never produces collisions.

Consider the following task for an ``adversary'' \(\adv\). They are given a function \(f\) which may be either a RP or a RF, and their goal is to determine which one they are interacting with by querying $f$ up to $Q$ times and observing the results, \ie, they are not allowed to, say, inspect the code of \(f\). Concretely, \(\adv\) should return \(\True\) if it interacts with a RP and \(\False\) for a RF.

How can \(\adv\) distinguish the two?
If \(\adv\) finds a collision, then it knows that $f$ cannot be a RP. 
However, if the domain of $f$ is very large compared to $Q$, then \(\adv\) cannot simply search the entire domain for a collision, and its chances of finding a collision are low.
Thus, the adversary will have a low chance of correctly distinguishing the two scenarios.
The switching lemma makes this formal by showing that the probability that \(\adv\) returns \(\True\) for either interaction differs by at most \(\frac{Q(Q-1)}{2\card{\dom{f}}}\).

\begin{lemma}[PRP/PRF Switching Lemma]\label{lem:switching}
  Let \(\adv\) be an adversary that asks at most \(Q\) queries and let \(N = \card{\dom\,\PRF} = \card{\dom\,\PRP}\). Then
  \[
    \abs{ \pr{\adv(\PRP) = \True} - \pr{\adv(\PRF)  = \True} } \leq \frac {Q (Q-1)} {2 N}\,.
  \]
\end{lemma}
The Switching Lemma gained notoriety because several published proofs of the lemma were found to contain mistakes~\citep{PRPPRF2}.
This observation was among the motivations for the development of a rigorous framework for cryptographic proofs such as the ones based on ``games'' \cite{PRPPRF2,Shoup:Sequences:2004} and the subsequent development of mechanized tools for such proofs \cite{cryptoverif,pRHL,easycrypt}.

To simplify the exposition, we first prove in \theaplog an instantiation of the lemma with a concrete \emph{weak adversary} $\advw$, which picks the input for its $Q$ queries \(x_0,\ldots,x_{Q-1}\) uniformly at random, without adapting to the response of the queries, and returns the list of outputs:
\[
  \begin{tabular}{>{$}l<{$} @{ } >{$}c<{$} @{\; } >{$}l<{$} }
    \DumbLet \advw\ \tapebound\ Q\ f
    =\\
    \quad\quad \Let \vxys = \Alloc \listempty in\\
    \quad\quad \For i = 0 to (Q-1) do \\
    \quad\quad\quad \Let \vx = \Rand~(\tapebound - 1) in \\
    \quad\quad\quad \Let \vy = f\, \vx in \\
    \quad\quad\quad \vxys \gets (\vx,\vy) :: \deref \vxys \\
    \quad\quad \deref \vxys
  \end{tabular}
\]

\begin{lemma}[Weak PRP/PRF Switching Lemma]\label{lem:wsl}
	Let $\advw(\tapebound, Q, \cdot)$ be the weak adversary defined above,
	and let \(N = \card{\dom\,\PRF} = \card{\dom\,\PRP}\). Then, for any list of results \(\vec{xy}\)
  \[
    \abs{ \pr{\advw(\tapebound, Q, \PRP) = \vec{xy}} - \pr{ \advw(\tapebound, Q, \PRF)  = \vec{xy}} } \leq \frac {Q (Q-1)} {2 N}\,.
  \]
\end{lemma}
We will prove the full Switching Lemma for an arbitrary \(Q\)-query adversary later in \cref{sec:switching-lemma}.

\paragraph{Random Functions and Permutations in \thelang}

First we need to model random functions and permutations as programs in \thelang.
An example implementation of an idealized random function with domain \(X = \{0,\ldots,\tapebound - 1\}\) is given by \(\erf\) in \cref{fig:prf-prp-alt}.
Upon initialization, \(\erf\,N\) creates a reference to an initially empty (finite) map \(\vm\) and returns a function \(\rf\).
On every call to \(\rf(x)\), if \(\rf\) has never been evaluated before on \(x\), a new point \(y \in X\) is sampled at random and stored into \(\vm\). Conversely, if \(x\) has been queried before, \(\rf(x)\) looks up its value in \(\vm\).

The idealized random permutation \(\erp\) likewise initializes its internal state and returns a closure~\(\rp\). However, to sample a new element, \(\rp\) randomly picks (the index \(k\) of) an element \(y\) of the list \(\vunused\) of values in \(X\) that do not yet occur in the codomain of \(\rp\),
removes \(y\) from \(\vunused\), and updates its internal map.
Initially, all values in \(X\) are unused, but as \(\rp\) is evaluated on new %
points, %
\(\vunused\) shrinks.
Since each element of \(X\) occurs only once in \(\vunused\) and gets removed the first time it is picked, \(\rp\) is guaranteed to remain collision-free.

\begin{figure}
  \begin{minipage}[t]{0.39\linewidth}\hspace{-1ex}
    \begin{tabular}[t]{>{$}l<{$} @{} >{$}c<{$} @{} >{$}l<{$} }
      \DumbLet \erf&&\,\, \tapebound = \\
                   && \DumbLet \vm = \mapinit\, \TT \In \\
      \\
                   && \Fun \vx . \If \mapget\ \vm\ \vx = \None then \\
                   && \quad\qquad \Let \vy = \Rand\, (\tapebound - 1) in \\
                   && \quad\qquad \mapset\ (\deref \vm)\ \vx\ \vy \Seq\\
      \\
      \\
      \\
                   && \quad\quad \mapget\ \vm\ \vx
    \end{tabular}
  \end{minipage}
  \quad
  \begin{minipage}[t]{0.54\linewidth}\hspace{-1em}
    \begin{tabular}[t]{>{$}l<{$} @{} >{$}c<{$} @{} >{$}l<{$} }
      \DumbLet \erp&&\,\, \tapebound = \\
                   &&\DumbLet \vm = \mapinit\, \TT \In \\
                   & & \Let \vunused = \Alloc~(\langv{List.seq}\ 0\ \tapebound) in
      \\
                   &&
                      \Fun \vx . \If \mapget\ (\deref \vm)\ \vx = \None then \\
                   &&\quad\qquad \Let \vlen = \listlen\ (\deref \vunused) in \\
                   &&\quad\qquad \Let \vk = \Rand\, (\vlen -1) in \\
                   &&\quad\qquad \Let \vy = \listnth\ (\deref \vunused)\ \vk in \\
                   &&\quad\qquad  \mapset\ (\deref \vm)\ \vx\ \vy \Seq\\
                   &&\quad\qquad  \vunused \gets \listremoventh\ (\deref \vunused)\ \vy \Seq\\
                   &&\qquad \mapget\ \vm\ \vx
    \end{tabular}
  \end{minipage}
  \caption{Example implementation of idealized RF and RP, parameterized by \(\tapebound = |\dom\,\erf| = |\dom\,\erp|\). }
  \label{fig:prf-prp-alt}
\end{figure}

\paragraph{The Mathematical Intuition}
We first give an informal sketch of why the result holds---the formal argument in \theaplog will closely mirror this style of reasoning.
We want to show that with error probability at most \(\err\), \(\advw(\tapebound, Q, \erp~\tapebound)\) and \(\advw(\tapebound, Q, \erf~\tapebound)\) compute the same list of results \(\vxys\), where \(\err\) is the bound from \cref{lem:wsl}.
After initializing the weak adversary, both lists are empty.
We claim that the lists remain equal with high probability through each iteration of the \(\langkw{for}\) loop.
W.l.o.g, we can assume both random samplings of \(\vx\) return the same value. If \(\vx\) has been sampled before, then no new information is gained from the call to \(f\), and the results are equal with the same probability as before.
If \(\vx\) is fresh, then \(\erf\) will sample a new response \(\vy\) out of \(\{0,\ldots,\tapebound - 1\}\), while \(\erp\) picks an element from \(\vunused\).
On the \(i\)-th loop iteration, \(\vunused\) contains at least \(\tapebound-i\) elements, and hence the probability of \(\erf\) sampling an element that \emph{does not} occur in \(\vunused\) and hence causes an observable collision is \(i/\tapebound\).
If \(\erf\) remains collision free, the probability that both programs compute the same result \(\vxys\) does not change.
We can hence establish an upper bound on the probability that \(\advw(\tapebound, Q, \erp~\tapebound)\) and \(\advw(\tapebound, Q, \erf~\tapebound)\) produce different results by summing the probabilities that each loop iteration observes a collision.
Since \(\sum_{i=0}^{Q-1} i/\tapebound = \frac {Q(Q-1)}{2 \tapebound}\), \cref{lem:wsl} holds.

\paragraph{The Proof in \Theaplog}
We derive \cref{lem:wsl} by proving the following pair of refinements.
\begin{lemma}\label{lem:wsl-refinement}
  Let \(Q, \tapebound \in \nat\), and let \(\err_Q \triangleq \frac {Q(Q-1)} {2 \tapebound}\). Then
  \begin{align}
    \upto{\err_Q}
    &\vdash
    \wpref{\advw(\tapebound, q, \erp~\tapebound)} {\advw(\tapebound, Q, \erf~\tapebound)}
    { x, y \ldotp x = y}\,\text{, and} \label{eq:wsl-prp-prf}\\
    \upto{\err_Q}
    &\vdash
    \wpref{\advw(\tapebound, Q, \erf~\tapebound)} {\advw(\tapebound, Q, \erp~\tapebound)}
    { x, y \ldotp x = y}\,. \label{eq:wsl-prf-prp}
  \end{align}
\end{lemma}
We sketch the proof of \eqref{eq:wsl-prp-prf}; the other direction is analogous.
We prove \eqref{eq:wsl-prp-prf} by reasoning backwards from the conclusion.
As a first step, we symbolically evaluate \(\advw(\tapebound, q, \erp~\tapebound)\). Evaluation order forces $\erp~\tapebound$ to evaluate first, which allocates a map \(\vm\) and the list \(\vunused\), and returns a function \(\rp\).
Afterwards, $\advw$ allocates a list of results \(\vxys\).
Similarly, evaluating \(\advw(\tapebound, q, \erf~\tapebound)\) allocates \(\vm'\), substitutes \(\rf\) for \(f\), and then allocates \(\vxys'\). Note that we will often prove refinements of the style \(\wpref {\expr_1}{\expr_2} \Phi\) where both \(\expr_1\) and \(\expr_2\) manipulate a variable \(x\). We frequently write \(x\) and \(x'\) for the left-hand side (\(\expr_1\)) and right-hand side (\(\expr_2\)) version of \(x\). Both \(\vxys\) and \(\vxys'\) point to the empty list at this stage, and the maps \(vm\) and \(vm'\) are empty.
We use the traditional ``points-to'' connective $p \mapsto v$ from separation logic to say that in the left program, $p$ points to $v$, and write $p \spointsto v$ for the analogous fact about the right program state.
We are thus left to prove the following refinement
\begin{equation*}
  \begin{array}[c]{l}
    \upto{\err_Q}
    \sep \vxys \pointsto []
    \sep \vxys' \spointsto [] \\[1mm]
    \sep \isprp\ \emptyset\ [0,\ldots,\tapebound]
    \sep \isprf\ \emptyset
  \end{array}
  \vdash \wpref
  {(loop_{\rp}\ Q\,\Seq \deref \vxys)}
  {(loop_{\rf}\ Q\,\Seq \deref \vxys')}
  {x,y \ldotp x=y}\,,
\end{equation*}
where \(loop_f\ Q\) stands for the \(\langkw{for}\) loop with bound \(Q\), and $\rp$ and $\rf$ are the functions returned by initializing \(\erp\) and \(\erf\) respectively.
The proposition \(\isprp\ m\ l\) means that \(\vm\) currently points to the map \(m\) and \(\vunused\) points to the list \(l\), and \(\isprf\ m\) likewise means that \(\vm'\) tracks the map \(m\).

We can generalize the goal slightly, and instead show the refinement below. The proof goes by induction on the number $i$ of remaining loop iterations (initially \(Q\)):
\begin{gather}\label{eq:wsl-ind}
    \upto {\err_i}
    \sep
    (\Exists m, l .
    \isprp\ m\ l
    \sep \vlen_i \leq \card{l}
    \sep \isprf\ m)
    \sep \Phi_{\mathit{res}}
    \wand
    \wpref {(loop_{\rp}\ i)} {(loop_{\rf}\ i)} {\Phi_{\mathit{res}}}\\[1mm]
    \text{where\;}
    \err_i \eqdef{} \textstyle\sum_{k=Q-i}^{Q-1}\frac k \tapebound = \frac {i(2Q - i - 1)} {2 \tapebound}
    \;,\;\,
    \vlen_i \eqdef{} \tapebound - (Q - i)
    \;,\;\,
     \Phi_{\mathit{res}} \eqdef{} (\Exists \vec{xy}.\vxys \pointsto \vec{xy} \sep \vxys' \spointsto \vec{xy})\,. \nonumber
\end{gather}
This maintains the key loop invariant that both $rp$ and $rf$ currently have the same mapping $m$.
The base case \(i=0\) holds, since the loop terminates immediately and the postcondition \(\Phi_{\mathit{res}}\) holds by assumption. To show the inductive case \(i=j+1\), we have to (1) prove that unrolling the loop once preserves \(\Phi_{\mathit{res}}\) and then (2) apply the induction hypothesis to the remaining \(j\) iterations.

By assumption, we start with \(\upto{\err_{j+1}}\) error credits, which represent the ``budget'' we can spend on avoiding collisions caused by calls to \(\rf\) that would result in different results.
From an easy calculation if follows that $\err_{j+1} = \frac{Q-i}{\tapebound} + \err_{j}$.
Our first step is to split this budget resource into two parts $\upto{\frac{Q-i}{\tapebound}} \sep \upto{\err_{j}}$.
We will use \(\upto{\frac{Q-i}{\tapebound}}\) to ``pay'' for avoiding collisions in the current loop iteration and \(\upto{\err_j}\) to account for the remaining \(j\) iterations.

We now focus on the first loop unrolling.
The first instruction in the loop body samples an input $x$ from \(\{0,\ldots,\tapebound - 1\}\) that will be used as the next query.
Since both programs sample from the same uniform distribution $\Rand\ (\tapebound - 1)$, we can use the exact coupling rule \ruleref{wp-couple-exact} seen earlier to proceed under the assumption that both programs sample the same value \(n\).
Next, the programs call \(\rp\ n\) and \(\rf\ n\) respectively. If \(n\) has been seen before, both functions return the same value.
If \(n\) is fresh on the other hand, \(\rf\) samples \(\vy\) from \(\{0,\ldots,\tapebound - 1\}\) whereas \(\rp\) picks an element from the list of unused values \(l\).
By assumption, the length of \(l\) is (at least) \(\vlen_i\).
The probability that \(\rf\) produce a collision, \ie, sample an element that is \emph{not} in \(l\), is \(\frac{\tapebound-\vlen_i}{\tapebound} = \frac{Q-i}{\tapebound}\).
Since this matches the error credits we have, we can apply the following \emph{approximate} coupling rule of \theaplog, which generalizes the simpler version seen earlier:
\begin{equation*}
  \infrule*[lab]{wp-couple-rand-rand-err-le}
  {
    g: \nat_{\leq K} \rightarrow \nat_{\leq \tapeboundB}~\text{injection} \\
    \upto{\tfrac{\tapeboundB-K}{\tapeboundB+1}}\\
    K \leq \tapeboundB\\
    \All k \leq K .
    \wpref{k}{
      g(k)
    }{\Phi}
  }
  {\wpref{\Rand K}{\Rand \tapeboundB}{\Phi}}
\end{equation*}
We instantiate \(g\) with \((\Lam n . \listnth\ l\ n )\), instantiate \(\tapeboundB\) with \(\tapebound - 1\), and instantiate \(K\) with \(\vlen_i -1 = \tapebound - (Q-i)-1\).
By giving up ownership of \(\upto{\frac{Q-i}{\tapebound}}\) for the second premise, we can thus continue the proof under the assumption that the \(\Rand\, (\vlen-1)\) and \(\Rand\, (\tapebound-1)\) resolve to a pair of values \(k\) and \(g(k)\) such that \(g(k)\), the newly sampled element in \(\rf\), is exactly the \(k\)-th element of \(l\).
Since we assumed \(\Phi_{\mathit{res}}\) as a hypothesis in \cref{eq:wsl-ind}, and since both programs add the same value \((n, g(k))\) to \(\vxys\) and  \(\vxys'\) respectively,  \(\Phi_{\mathit{res}}\) holds again after the first loop unfolding.
We can thus conclude our proof by appealing to the induction hypothesis, paying for the error credit premise with \(\upto{\err_j}\).

\subsection{Error Amplification for Exact Equivalences}
\label{sec:key-ideas-rej-samp}
\newcommand{\rejsampler}{\langv{sampler}}
As alluded to in the introduction, error credits not only allow us to prove approximate equivalences, they also allow us to prove \emph{exact} equivalences of probabilistic programs.
To motivate this, consider the following rejection sampler, where $\tapeboundB < \tapebound$.
\begin{align*}
  &\Rec \rejsampler ~\_ =  \Let x = \Rand\tapebound in  \If x \leq \tapeboundB then x \Else \rejsampler~()
\end{align*}
By continuously rejecting samples which do \emph{not} correspond to values in the  target set $\{0, \ldots, \tapeboundB\}$ and retrying, $\langv{sampler}$ eventually produces values that are sampled uniformly from $\{0, \dots, \tapeboundB\}$.
We can state this formally by proving that $\langv{sampler}$ is equivalent to the expression $\Rand\tapeboundB$, \ie, by showing the following two refinements:
\begin{align*}
  &\wpref{\rejsampler~()}{\Rand\tapeboundB}{\val_{1}, \val_{2} \ldotp \val_{1} = \val_{2}},~\text{and} \\
  &\wpref{\Rand\tapeboundB}{\rejsampler~()}{\val_{1}, \val_{2} \ldotp \val_{1} = \val_{2}}.
\end{align*}
Perhaps surprisingly, although this kind of equivalence is a standard and important result in randomized algorithms, no existing relational program logic can establish this with couplings, to the best of our knowledge, even with approximate couplings.

To see what goes wrong, let us focus on trying to prove the second refinement, and consider trying to apply a coupling rule to the step where the left program executes $\Rand \tapeboundB$ and the right executes $\Rand \tapebound$.
If the right program's sample is $\leq \tapeboundB$, then we want that value to be coupled and equal to the value sampled on the left.
On the other hand, if the right sample is $> \tapeboundB$, then it will be rejected, so we do not want to couple the result of $\Rand \tapeboundB$ on the left to this value at all.
We only want to couple the left sample to be equal to the eventual later value that ends up being accepted!
Trying to use an approximate coupling to force both samples to be equal will not work either, as that would incur a large error, and we are trying to show an exact coupling.

\theaplog overcomes this limitation by introducing a new form of couplings called \emph{fragmented} couplings, which can be combined with a technique called \emph{error amplification} introduced by Eris.
To start, rather than proving the above refinements with no error credits, we instead merely have to prove them starting with an arbitrarily small positive error credit. That is, we must show:
\begin{align*}
  \upto{\err} \vdash \wpref{\Rand\tapeboundB}{\rejsampler~()}{\val_{1}, \val_{2} \ldotp \val_{1} = \val_{2}}
\end{align*}
for \emph{all} $\err > 0$.
The exact refinement then follows by a limiting argument, as $\err \rightarrow 0$, \cf{}, \cref{thm:lim-ade}.

But what can we do with an arbitrarily small error credit?
After all, it may be too small to apply the intended approximate coupling rule.
The solution is that when reasoning about a $\Rand N$ command, the logic allows us to \emph{amplify} and grow the $\err$ credits along branches of the random outcome, as long as the expected amount of error credit across all branches is still $\err$. In particular, the following \emph{expectation-preserving} fragmented coupling will allow us to apply this principle.
\begin{mathpar}
  \inferrule*
  {
    \tapeboundB<\tapebound \\
    \upto{\err} \\
    \ghostcode{\progtape{\lbl}{\tapeboundB}{\err}} \\
    \\
    \All m \leq \tapeboundB . \wpref{m}{m}{\Phi} \\
    \All m > \tapeboundB .
    \ghostcode{
      \progtape{\lbl}{\tapeboundB}{\err} \sep
    }
    \upto{\tfrac{\tapebound+1}{\tapebound-\tapeboundB}\cdot \err } \wand \wpref{\Rand~\tapeboundB~\ghostcode{\lbl}}{m}{\Phi}
  }
  { \wpref{\Rand~\tapeboundB~\ghostcode{\lbl}}{\Rand~\tapebound}{\Phi} }
\end{mathpar}
Here we present a simplified variant of a more general rule, and the \ghostcode{gray}ed out parts may be ignored for now.
We discuss the general rules in \cref{sec:rules}.
The rule requires $\err$ credits and lets us relate two sampling operations, $\Rand\tapeboundB$ and $\Rand\tapebound$.
It asks us to consider two cases: (1) the outcome of the two samplings agree and are within range, and (2) the right sampling is resolved to some $m > \tapeboundB$ but the left does not sample anything; we also get to assume ownership of $\tfrac{\tapebound+1}{\tapebound-\tapeboundB}\cdot \err$ error credits.
We call this a ``fragmented'' coupling because it allows for one of the coupled programs to not necessarily execute its random sample, depending on what the other program drew.

Applying this rule to the rejection sampler, when the first case occurs, the sample will be accepted and the proof will conclude.
In the second case, the rejection sampler will loop.
Now, for any starting $\err$, if we repeatedly increase our error credits by a factor of $\tfrac{\tapebound+1}{\tapebound-\tapeboundB}$ by each loop iteration, then eventually we will have a large enough error credit to apply an approximate coupling rule and ``force'' the right-hand sample to be in range.
By doing induction on the number of amplifications needed, we can therefore conclude the proof.

%% file: preliminaries.tex
\section{Preliminaries}
\label{sec:preliminaries}

In this section, we recall some basic definitions in probability theory and a notion of \textit{approximate couplings} \cite{Sato:Approximate:2016}.
We then introduce the syntax and semantics of \thelang, the language of our programs, and our notion of contextual equivalence.

\subsection{Probability Theory}
\label{sec:prelim-prob}

To account for possibly non-terminating behavior of programs, we define our operational semantics using probability \emph{sub}-distributions.
\begin{definition}[Distribution]
  A \defemph{discrete subdistribution} (henceforth simply \defemph{distribution}) on a countable set $A$ is a function $\distr : A \ra [0,1]$ such that $\sum_{a \in A}\distr(a) \leq 1$.
  The collection of distributions on $A$ is denoted by $\Distr A$.
\end{definition}
Given a predicate $P$, the \emph{Iverson bracket} $\iverbr{P}$ evaluates to $1$ if $P$ is true and to $0$ otherwise.
\begin{lemma}[Discrete Distribution Monad]
	We can equip $\DDistr$ with a monadic structure, with operations
	\begin{align*}
		&{}\mret \colon A \to \Distr A \qquad &{}\mbind\colon (A \to \Distr B) \to \Distr A \to \Distr B \\
		&{}\mret(a)(a') \eqdef{} \iverbr{a = a'} &{}\mbind(f,\distr)(b) \eqdef{} \sum_{a \in A} \distr(a) \cdot f(a)(b)
	\end{align*}
  We use the notation $\distr \mbindi f$ for $\mbind(f, \distr)$.
\end{lemma}
\begin{definition}[Expected value]
  Let $\distr \in \Distr{A}$ be a distribution and $X \colon A \to [0,1]$ a random variable. The \emph{expected value} of $X$ with respect to $\distr$ is defined as
        $  \expect[\mu]{X} \eqdef{} \sum_{a \in A} \mu(a) \cdot X(a) $.
\end{definition}
Many probabilistic relational program logics use \emph{probabilistic couplings} \cite{thorisson/2000, lindvall_lectures_2002, Villani2008OptimalTO}, a mathematical tool for reasoning about pairs of probabilistic processes.
To reason about approximate equivalence of probabilistic programs, we use a notion of \emph{approximate} probabilistic coupling \cite{Sato:Approximate:2016}.
\begin{definition}[Approximate Coupling]
  Let $\distr_1 \in \Distr{A}$ and $\distr_2 \in \Distr{B}$.
  Given some approximation error $\err \in [0,1]$ and a relation $R \subseteq A \times B$, we say that there exists an $(\err, R)$-coupling of $\distr_1$ and $\distr_2$ if for all $[0,1]$-valued random variables $X : A \ra [0,1]$ and $Y : B \ra [0,1]$, such that $(a,b) \in R$ implies $X(a)\leq Y(b)$, the expected value of $X$ exceeds the expected value of $Y$ by at most $\err$, \ie, $ \expect[\distr_1]{X} \leq \expect[\distr_2]{Y} + \err$.
  We write \(\ARcoupl {\distr_1} {\distr_2} \err R\) if an $(\err,R)$-coupling exists between \(\distr_1\) and \(\distr_2\).
\end{definition}

Proving existence of $(\err,R)$-couplings for particular choices of $R$ is useful to prove relations between distributions.
When $R$ is the equality relation, couplings can be used to prove bounds on the total variation distance, which has applications when reasoning about convergence properties, as well as in security definitions.

\begin{lemma}\label{lem:arc-elim}
  Let \(\distr_1, \distr_2 \in \Distr A\) such that there exists an \(\err\)-coupling for the equality relation, \ie, \(\ARcoupl {\distr_1} {\distr_2} \err (=)\), then for all  \( a \in A\) we have \(\distr_1(a) \leq \distr_2(a) + \err\).
  If, in addition, \(\ARcoupl {\distr_2} {\distr_1} \err (=)\), then the total variation distance between \(\distr_1\) and \(\distr_2\) is at most \(\err\), \ie, \(\sup_{S \subseteq A} | \distr_1(S) - \distr_2(S) | \leq \err\).
\end{lemma}

\begin{corollary}\label{cor:arcoupl-elim}
  Let \(\distr_1, \distr_2 \in \Distr A\).
  Then \(\ARcoupl {\distr_1} {\distr_2} 0 (=)\) %
  implies that for all \(a \in A \), \( \distr_1(a) \leq \distr_2(a)\).
\end{corollary}
By completeness of the real numbers, we obtain the following limiting theorem.
\begin{lemma}\label{lem:arcoupl-lim}
  Let \(\distr_1, \distr_2 \in \Distr A\) and $\err \in [0,1]$.
  If \(\ARcoupl {\distr_1} {\distr_2} {\err'} R\) for all $\err' > \err$ then \(\ARcoupl {\distr_1} {\distr_2} {\err} R\).
\end{lemma}

To construct couplings between program executions, we can compose couplings of single steps of executions.
This is possible because couplings compose along the bind of the distribution monad.
Let $\distr_1 \in \Distr A$, $\distr_2 \in \Distr B$, $f : A \ra \Distr {A'}$, $g : B \ra \Distr {B'}$, $R \subseteq A \times B$, and \(R' \subseteq A' \times B'\).
\begin{lemma}
  If \(\ARcoupl {\distr_1} {\distr_2} \err R\) and
  \(\forall (a,b) \in R, \ARcoupl {f(a)} {g(b)} {\err'} {R'}\),
  then \(\ARcoupl {(\distr_1 \mbindi f)} {(\distr_2 \mbindi g)} {\err + \err'} {R'}\).
\end{lemma}

We can strengthen this lemma by letting the grading \(\err'\) for the continuation vary depending on the value that \(a\) takes, and consider its expected value \wrt \(\distr_1\) when composing the couplings:
\begin{lemma}\label{lem:arcoupl-exp-l}
  Let \(\Err : A \ra [0,1]\).
  If \(\ARcoupl {\distr_1} {\distr_2} \err R\)
  and \(\forall (a,b) \in R, \ARcoupl {f(a)} {g(b)} {\Err(a)} {R'}\),
        then \(\ARcoupl {(\distr_1 \mbindi f)} {(\distr_2 \mbindi g)} {\err + \err'} {R'}\) where \(\err' = \expect[\distr_1]{\Err}\).
\end{lemma}
Symmetrically, we can vary the error on \(B\) and consider its expected value \wrt \(\distr_2\):
\begin{lemma}\label{lem:arcoupl-exp-r}
  Let \(\Err : B \ra [0,1]\).
  If \(\ARcoupl {\distr_1} {\distr_2} \err R\)
  and \(\forall (a,b) \in R, \ARcoupl {f(a)} {g(b)} {\Err(b)} {R'}\),
  then \(\ARcoupl {(\distr_1 \mbindi f)} {(\distr_2 \mbindi g)} {\err + \err'} {R'}\)
  where \(\err' = \expect[\distr_2]{\Err}\).
\end{lemma}
To the best of our knowledge, the \defemph{expectation-preserving composition lemmas} \labelcref{lem:arcoupl-exp-l,lem:arcoupl-exp-r} are novel, at least in the context of program logics. We apply these results for rules such as the expectation-preserving fragmented coupling rule presented in \cref{sec:key-ideas-rej-samp}, where we can amplify the error credit for certain branches as long as the expected amount of error credit across all branches remains the same.

\subsection{The \thelang Language and Operational Semantics}
\label{sec:prelim-lang}

The $\thelang{}$ language that we consider is an ML-like language with probabilistic uniform sampling, higher-order functions, higher-order state, recursive types, and impredicative type polymorphism.

The syntax is defined by the grammar below.
\begin{align*}
  \val, \valB \in \Val \bnfdef{}
  & z \in \integer \ALT
  b \in \bool \ALT
  \TT \ALT
  \loc \in \Loc \ALT
  \Rec \vf \lvar = \expr \ALT
  (\val,\valB) \ALT
  \Inl \val  \ALT
  \Inr \val
  \\
  \expr \in \Expr \bnfdef{}  &
  \val \ALT
  \lvar \ALT
  \Rec \vf \lvar = \expr \ALT
  \expr_1~\expr_2 \ALT
  \expr_1 + \expr_2 \ALT
  \expr_1 - \expr_2 \ALT
  \ldots \ALT
  \If \expr then \expr_1 \Else \expr_2 \ALT
  (\expr_1,\expr_2) \ALT
  \Fst \expr \ALT \ldots \\
  & \Alloc~\expr_1 \ALT
  \deref \expr \ALT
  \expr_1 \gets \expr_2 \ALT
  \expr_1 [\expr_2] \ALT
  \Rand~\expr \ALT
  \Pack \expr \ALT
  \Unpack \expr as \var in \expr \ALT \ldots \\
  \lctx \in \Ectx \bnfdef{}  &
  -
  \ALT \expr\,\lctx
  \ALT \lctx\,\val
  \ALT \Alloc~\lctx
  \ALT \deref \lctx
  \ALT \expr \gets \lctx
  \ALT \lctx \gets \val
  \ALT \Rand \lctx
  \ALT \ldots
  \\
  \state \in \State \eqdef{} & \Loc \fpfn \Val \hspace{2em} \cfg \in \Conf \eqdef{} \Expr \times \State \\
  \type \in \Type \bnfdef{}
  & \alpha \ALT
    \tunit \ALT
    \tbool \ALT
    \tnat \ALT
    \tint \ALT
    \type \times \type \ALT
    \type + \type \ALT
    \type \to \type \ALT
    \All \alpha . \type \ALT
    \Exists \alpha . \type \ALT
    \tmu \alpha . \type \ALT
    \tref{\type}
\end{align*}
The term language is mostly standard.
We use $\Alloc \expr_1$ to allocate a new reference containing the value returned by $\expr_1$,
$\deref \expr$ to dereference the location $\expr$ evaluates to, and $\expr_{1} \gets \expr_{2}$ to evaluate $\expr_{2}$ and assign the result to the location that $\expr_{1}$ evaluates to.
We often refer to a recursive function value $\Rec \vf \lvar = \expr$ by its name $\vf$.
The operation $\Rand~\tapebound$ denotes uniform random sampling over $\{ 0, \ldots, \tapebound \}$.

Finally, we have several terms related to typing operations
\eg, $\Pack \expr$ and $\Unpack \expr_{1} as \var in \expr_{2}$ are used for introducing and eliminating existential types.
We write $\pfctx \mid \vctx \proves \expr \colon \type$ to denote that $\expr$ has type $\type$ in the typing context $\pfctx \mid \vctx$, which consists of a context of type variables $\pfctx$ and a context of program variables $\vctx$.
The inference rules for the typing judgments are standard (see, \eg, \citet{reloc} or the Coq formalization).%

\paragraph{Operational Semantics}

To define program execution, we define $\stepdistr(\cfg) \in \Distr{\Conf}$, the distribution induced by the single step reduction of configuration $\cfg \in \Conf$.
The semantics is mostly standard.
We first define head reductions and then lift it to reduction in an evaluation context $K$.
All non-probabilistic constructs reduce deterministically as usual, \eg, %
$\stepdistr((\Fun \vx . \expr)\,\val, \sigma) = \mret(\expr[\val/\vx], \sigma)$.
We write \(e \purestep e'\) if the evaluation is deterministic and holds independently of the state, \eg, \((\Fun \vx . \expr)\,\val \purestep \expr[\val/\vx]\) and \(\Fst (\val_1, \val_2) \purestep \val_1\).
The probabilistic choice $\Rand \tapebound$ reduces uniformly at random, \ie,
\begin{align*}
  & \stepdistr(\Rand \tapebound, \sigma)(n, \sigma) \eqdef{}
  \begin{cases}
    \frac{1}{\tapebound + 1} & \text{for } n \in \{ 0, 1, \ldots, N \}, \\
    0                        & \text{otherwise}.
  \end{cases}
\end{align*}

With the single step reduction $\stepdistr(-,-)$ defined, we next define a step-stratified execution probability $\exec_{n}\colon \Conf \to \Distr{\Val}$ by induction on $n$:
\begin{align*}
	\exec_{0}(\expr, \state)(\val) &\eqdef{}
 \begin{cases}
   1                                           & \text{if}~\expr\in\Val \wedge \expr = \val, \\
   0 & \text{otherwise.}
 \end{cases}\\
	\exec_{m+1}(\expr, \state)(\val) &\eqdef{}
 \begin{cases}
   1                                           & \text{if}~\expr\in\Val \wedge \expr = \val, \\
   \sum_{(\expr',\state')\in \Expr\times\State} \stepdistr(\expr, \state)(\expr',\state') \cdot \exec_{m}(\expr',\state')(\val) & \text{otherwise.}
 \end{cases}
\end{align*}
That is, $\exec_{n}(\expr, \state)(\val)$ is the probability of stepping from the configuration $(\expr, \state)$ to a value $\val$ in less than $n$ steps.
The probability that a execution, starting from configuration $\cfg$, reaches a value $\val$ is taken as the limit of its stratified approximations, which exists by monotonicity and boundedness:
\begin{align*}
 \exec(\cfg)(\val) \eqdef{} \lim_{n \to \infty} \exec_{n}(\cfg)(\val)
\end{align*}
The termination probability of an execution from configuration $\cfg$ is $\execTerm{(\cfg)} \eqdef{} \Sigma_{\val\in\Val} \exec(\cfg)(v)$.

The definition of program execution as a distribution leads to a natural notion of $\err$-approximation.
We say that \(\expr_1\) \defemph{\(\err\)-approximates} \(\expr_2\) if
$
  \exec (\expr_1,\state)(\val) \leq \exec(\expr_2,\state)(\val)+\err
$
for all $\val,\state$.
By \cref{lem:arc-elim}, we can show such approximations by establishing an approximate coupling of the executions of \(\expr_1\) and \(\expr_2\).
We say \(\expr_1\) and \(\expr_2\) are \(\err\)-equivalent if both \(\expr_1\) \(\err\)-approximates \(\expr_2\) and \(\expr_2\) \(\err\)-approximates~\(\expr_1\). In that case, we have \(\abs{\exec(\expr_1,\state)(\val) - \exec(\expr_2,\state)(\val)} \leq \err\) for all $\val,\state$.

\begin{example}
 Consider the program below
	\[ e \eqdef{}  \Let x = \Rand\tapebound in  x \leq \tapeboundB \]
 and assume $\tapeboundB \leq \tapebound$. Evaluation order dictates that the random sampling is resolved first. Then we have,
 for any $\state\colon\State$ and any $0 \leq n \leq \tapebound$:
	\[ \stepdistr(e,\sigma)(\Let x = n in  x \leq \tapeboundB,\sigma) = \frac{1}{\tapebound+1} \]
 The probability of stepping to any other configuration is 0. Fixing a particular $n$, the next step is deterministic, and therefore we have
	\[ \stepdistr(\Let x = n in  x \leq \tapeboundB,\sigma) = \mret(n \leq \tapeboundB,\sigma)\]
 The final step is also deterministic and just evaluates the inequality $n \leq \tapeboundB$ to
 either $\True$ or $\False$. Collecting all of the probabilities of the succesful comparisons together, we can show
	\[ \exec_{3}(e,\sigma)(\True) = \frac{\tapeboundB+1}{\tapebound+1} \]
 and trivially at the limit
	\[ \exec(e,\sigma)(\True) = \frac{\tapeboundB+1}{\tapebound+1} \]
 Note that, since the execution of $e$ takes exactly 3 steps to reach to a value, executing \(e\) for fewer steps returns the zero distribution on values:
	\[ \exec_{0}(e,\sigma) = \exec_{1}(e,\sigma) = \exec_{2}(e,\sigma) = \lambda v. 0 \]
\end{example}

\paragraph{Presampling Tapes}
Standard probabilistic coupling logics require aligning or ``synchronizing''  sampling statements of the two programs under consideration.
For example, both programs have to be executing the sample statements we want to couple for their next step when applying a coupling rule.
However, it is not always possible to synchronize sampling statements in this way, especially for higher-order programs.
To address this issue, \citet{clutch} introduce \emph{asynchronous coupling}.
As we will see, the same mechanisms are useful for approximate relational reasoning.

Asynchronous couplings are introduced through dynamically-allocated \emph{presampling tapes} that are added to the language.
Intuitively, presampling tapes will allow us \emph{in the logic} to presample (and in turn couple) the outcome of future sampling statements.
Formally, presampling tapes appear as two new constructs added to the programming language.

\begin{minipage}{0.5\linewidth}
  \begin{align*}
  \val \in \Val \bnfdef{}& \ldots \ALT \lbl \in \Lbl \\
  \expr \in \Expr \bnfdef{}& \ldots \ALT
                             \AllocTape\,\expr \ALT
                             \Rand \expr_{1}~\expr_{2} \\
  \lctx \in \Ectx \bnfdef{}  & \ldots \ALT
                             \AllocTape\,\lctx \ALT
                             \Rand \expr~\lctx \ALT
                             \Rand \lctx~\val \\
\end{align*}
\end{minipage}
\hfill
\begin{minipage}{0.5\linewidth}
  \begin{align*}
  \state \in \State \eqdef{}& (\Loc \fpfn \Val) \times (\Lbl \fpfn \Tape) \\
    t \in \Tape \eqdef{}& \{ (\tapebound, \tape) \mid \tapebound \in \mathbb{N} \wedge \tape \in \mathbb{N}_{\leq \tapebound}^{\ast} \} \\
    \type \in \Type \bnfdef{}&  \ldots \ALT \ttape \\
  \end{align*}
\end{minipage}

\noindent The $\AllocTape \tapebound$ operation allocates a new fresh tape with label $\lbl$ and upper bound $\tapebound$, representing future outcomes of $\Rand \tapebound\, \lbl$ operations%
.
The $\Rand$ primitive can now (optionally) be annotated with the tape label $\lbl$.
If the corresponding tape is empty, $\Rand \tapebound\,\lbl$ reduces to any $n \leq \tapebound$ with equal probability, just as if it had not been labeled.
But if the tape is \emph{not} empty, then $\Rand\tapebound\,\lbl$ reduces \emph{deterministically} by taking off the first element of the tape and returning it.
\begin{align*}
	\stepdistr(\AllocTape \tapebound, \sigma) &\eqdef{}
	\begin{cases}
		\mret(\lbl, \sigma[\lbl\mapsto(\tapebound, \epsilon)]) \qquad &{}\lbl = {\sf fresh}(\sigma),\ \tapebound \geq 0\,, \\
		\mret(\lbl, \sigma[\lbl\mapsto(0, \epsilon)]) \qquad &{}\lbl = {\sf fresh}(\sigma),\ \tapebound < 0\,.
	\end{cases} \\
  \stepdistr(\Rand \tapebound\,\lbl, \sigma[\lbl\mapsto(\tapebound, \epsilon)])(n, \sigma[\lbl\mapsto(\tapebound, \epsilon)])
  &\eqdef{}
    \begin{cases}
      \frac{1}{\tapebound + 1} & \text{for } n \in \{ 0, 1, \ldots, N \}, \\
      0                        & \text{otherwise}.
    \end{cases} \\
  \stepdistr(\Rand \tapebound\,\lbl, \sigma[\lbl\mapsto(\tapebound, n::w)])
  &\eqdef{} \mret(n, \sigma[\lbl\mapsto(\tapebound, w)])
\end{align*}

Note that \emph{no} primitives in the language add values to the tapes.
Instead, values are added to tapes as part of presampling steps that will be \emph{ghost operations} appearing only in the logic.
In fact, labeled and unlabeled sampling operations are contextually equivalent \cite{clutch}.
This result follows from the fact that the ghost operations for adding values to tapes are \emph{erasable} in the following sense:

\begin{definition}[Erasable]
  \label{def:erasable}
  Let $\mu \in \Distr{\State}$ and $\state \in \State$.
  \[
    \erasable(\mu, \state) \eqdef{} \All \expr, n . \exec_{n} (\expr, \state) = \left(\mu \mbindi \Lam \state' . \exec_{n} (\expr, \state')\right)
  \]
\end{definition}
Erasability of $\mu$ w.r.t. $\state$ intuitively captures that distribution $\mu$ does not influence the probabilistic \emph{outcome} of any program execution from state $\sigma$.
For example, $\erasable(\mret(\state), \state)$ trivially holds by the left identity law of the distribution monad.
More interestingly, in \thelang{}, $\erasable(\statestepdistr{\lbl}(\state), \state)$ holds where $\statestepdistr{\lbl}(\state)$ is the distribution of the ghost operation that samples a fresh value uniformly onto the end of the presampling tape with label $\lbl$ in state $\state$.
This is the essence of the soundness of asynchronous couplings and ultimately what allows us to validate rules such as \ruleref{wp-tape-tape-append} and \ruleref{wp-many-to-one}, which we explain later in \cref{sec:rules}.

\subsection{Contextual Refinement and Equivalence}
\label{sec:prelim-equiv}
A program context $\ctx$ is an expression with a hole and we write $\fillctx\ctx[\expr]$ for the term resulting from replacing the hole in $\ctx$ by $\expr$.
Contexts are also typed; we write $(\ctx \colon (\pfctx \mid \vctx \proves \type) \Ra (\pfctx' \mid \vctx' \proves \type'))$ whenever $\pfctx' \mid \vctx' \proves \ctx[\expr] \colon \type'$ for every well-typed $\pfctx \mid \vctx \proves \expr \colon \type$.

The notion of contextual refinement that we use is standard and uses the termination probability $\execTerm$ as observation predicate.
We say expression $\expr_{1}$ \emph{contextually refines} expression $\expr_{2}$ if for all well-typed program contexts $\ctx$ resulting in a closed program then the termination probability of $\fillctx\ctx[\expr_{1}]$ is bounded by the termination probability of $\fillctx\ctx[\expr_{2}]$:
\begin{align*}
  \ctxrefines{\pfctx \mid \vctx}{\expr_{1}}{\expr_{2}}{\type} \eqdef{}
  &\All \type', (\ctx : (\pfctx \mid \vctx \proves \type) \Ra (\emptyset \mid \emptyset \proves \type')), \state .
  \, \execTerm(\fillctx \ctx [\expr_{1}], \state) \leq \execTerm(\fillctx \ctx [\expr_{2}], \state)
\end{align*}
Note that contextual refinement is a precongruence, and that the statement itself is in the meta-logic (\eg{}, Coq) and makes no mention of \theaplog{} or Iris.
We define \emph{contextual equivalence} $\ctxeq{\pfctx \mid \vctx}{\expr_{1}}{\expr_{2}}{\type}$ as refinement in both directions, \ie{}, $\ctxrefines{\pfctx \mid \vctx}{\expr_{1}}{\expr_{2}}{\type}$ and $\ctxrefines{\pfctx \mid \vctx}{\expr_{2}}{\expr_{1}}{\type}$.

%% file: logic.tex
\section{An Approximate Relational Logic}
\label{sec:arl}
In this section, we introduce the relational \theaplog logic and its soundness theorem, with an emphasis on the novel relational rules that interact with error credits or with presampling tapes.
We then demonstrate the logic on a simple rejection sampler.

\theaplog is built on top of the Iris separation logic framework~\cite{irisjournal} and hence inherits many of Iris's logical connectives.
A selection of \theaplog propositions is shown below.
\begin{align*}
  \prop,\propB \in \iProp \bnfdef{}
  & \TRUE \ALT \FALSE \ALT \prop \land \propB \ALT \prop \lor \propB \ALT \prop \Ra \propB \ALT
    \All \var . \prop \ALT \Exists \var . \prop \ALT \prop \sep \propB \ALT \prop \wand \propB \ALT \\
  & \progheap{\loc}{\val} \ALT
    \specheap{\loc}{\val} \ALT
    \progtape{\lbl}{\tapebound}{\tape} \ALT
    \spectape{\lbl}{\tapebound}{\tape} \ALT
    \upto{\err} \ALT
    \wpref{\expr_{1}}{\expr_{2}}{\val_{1}, \val_{2} \ldotp \prop} \ALT
    \ldots%
\end{align*}
Most of the propositions are standard, such as separating conjunction $\prop \sep \propB$ and separating implication $\prop \wand \propB$ (the magic wand).
As we saw earlier, the heap points-to assertion that denotes ownership of location $\loc$ comes in two forms: $\progheap{\loc}{\val}$ for the left-hand side, and $\specheap{\loc}{\val}$ for the right-hand side (the ``specification'' side).
Similarly, since presampling tapes are part of the state, we also have tape points-to assertions for both sides: $\progtape{\lbl}{\tapebound}{\tape}$ and $\spectape{\lbl}{\tapebound}{\tape}$, respectively.

Inspired by Eris~\cite{eris}, we interpret errors as resources in our logic using the $\upto{\err}$ connective for $\err\in[0,1]$.
Intuitively, $\upto{\err}$ denotes ownership of $\err$ error credits that can be spent to do $\err$-approximate reasoning.
Error credits can be split and combined, \ie{},  $\upto{\err_{1}+\err_{2}} \dashv\vdash \upto{\err_{1}}\sep\upto{\err_2}$.
Another important fact is that ownership of $1$ error credit immediately leads to a contradiction, \ie{}, $\upto{1}\vdash\FALSE$.
Intuitively, this is sound because there always exists a trivial $(1, \varphi)$-coupling for any two distributions and for any $\varphi$.

To show that $\expr_{1}$ $\err$-approximates $\expr_{2}$ we prove an entailment of the form $\upto{\err} \vdash \wpref{\expr_{1}}{\expr_{2}}{ v, v'. v=v' }$.
The following soundness theorem says we may then conclude the existence of an $\err$-approximate coupling under the equality relation.
\begin{theorem}[Adequacy]
  \label{thm:ade}
  Let $\varphi \subseteq \Val \times \Val$ be a relation and $\err \in [0,1]$. If
  \(
     \upto{\err} \vdash \wpref{\expr_{1}}{\expr_{2}}{\varphi}
  \)
  then $\ARcoupl{\exec(\expr_{1}, \state_{1})}{\exec(\expr_{2}, \state_{2})}{\err}{\varphi}$ for all $\state_{1}$ and $\state_{2}$.
\end{theorem}
The result is stated here to provide the reader with a semantic understanding of the rules we will present
in this section. In \cref{sec:model} we will explain the underlying model in more detail and discuss the
proof of this result.

As a corollary of the above, we get the following \emph{error-limiting} result by appealing to \cref{lem:arcoupl-lim}.
\begin{corollary}[Error-Limiting Adequacy]
  \label{thm:lim-ade}
  Let $\varphi \subseteq \Val \times \Val$ be a relation and $\err \in [0,1]$. If
  \(
     \upto{\err'} \vdash \wpref{\expr_{1}}{\expr_{2}}{\varphi}
  \)
  for all $\err' > \err$ then $\ARcoupl{\exec(\expr_{1}, \state_{1})}{\exec(\expr_{2}, \state_{2})}{\err}{\varphi}$ for all $\state_{1}$ and $\state_{2}$.
\end{corollary}

The corollary is similar to \cref{thm:ade} in that we obtain an approximate $(\err, \varphi)$-coupling of the execution of the two programs.
However, instead of establishing the weakest precondition given $\err$ credits, one has to prove the weakest precondition given $\err'$ credits for an arbitrary $\err' > \err$.
 Note that by picking $\err$ to be $0$ this also allows us to establish exact equivalences as we demonstrate in \cref{sec:rej-samp}.

\subsection{Rules of \theaplog}
\label{sec:rules}

In this section, we present a selection of the rules of \theaplog.
The rules are categorized into four classes.
We start with program-logic rules that are the relatively standard laws that most relational separation logics enjoy.
We then discuss of (approximate and non-approximate) coupling rules.
Afterwards, we consider rules that use presampling tapes to reason about more complicated couplings.
We conclude with a discussion of the error amplification proof technique which \theaplog supports for reasoning about recursive programs.

\paragraph{Program-Logic Rules}
We note that most rules (except where explicitly mentioned, in particular \ruleref{wp-rec}) have both left- and right-sided variants.
For brevity, we present only left-sided variants, right-sided variants are symmetric and use specification-side connectives ($\mapsto_{\mathsf{s}}$ and $\hookrightarrow_{\mathsf{s}}$).

Although \theaplog is a separation logic for reasoning about probabilistic programs, the rules of the non-probabilistic fragment are identical to the structural and computational rules found in most logics for non-probabilistic programs.
A selection of rules for the deterministic fragment is found in \cref{fig:deterministic-rules}.
For example, \theaplog satisfies the relational bind rule (\ruleref{wp-bind}), rules for symbolically taking deterministic ``pure'' steps---steps that do not depend on state (\ruleref{wp-pure-l}), and rules for interacting with the heap (\ruleref{wp-load-l}).
The rule \ruleref{wp-rec} is the standard recursive function rule found in general program logics.
(We do not have a ``standard'' recursive function rule for the right-hand side program because of how our refinement weakest-precondition assertion is defined. See \cref{sec:model} for more details.)

\begin{figure*}[th!]
  \centering
\begin{mathpar}
  \infrule{wp-bind}
  { \wpref{\expr_1}{\expr_2}{\Psi} \\ \All \val_1, \val_2 . \Psi(\val_1, \val_2) \wand \wpref{\fillctx\lctx[\val_1]}{\fillctx\lctx[\val_2]}{\Phi}}
  { \wpref{\lctx[\expr_1]}{\lctx'[\expr_2]}{\Phi}}
  \and
  \infrule{wp-pure-l}
  {
    \expr_1 \purestep \expr_1' \\
    \wpref{\expr_1'}{\expr_2}{\Phi}
  }
  {\wpref{\expr_1}{\expr_2}{\Phi}}
  \and
  \infrule{wp-load-l}
  {\progheap{\loc}{\val} \\
    \progheap{\loc}{\val} \wand \wpref{\val}{\expr}{\Phi} }
  {\wpref{\deref\loc}{\expr}{\Phi}} \\
  \infrule{wp-rec}
  { (\All \valB . \wpref{(\Rec f x = \expr)~\valB}{\expr'}{\Phi}) \vdash
    \wpref{\subst{\subst{\expr_1}{x}{\val}}{f}{(\Rec f x = \expr)}}{\expr'}{\Phi}
  }
  {\vdash \wpref{(\Rec f x = \expr)~\val}{\expr'}{\Phi} }
\end{mathpar}
\caption{A selection of the deterministic program-logic rules of \theaplog.}
\label{fig:deterministic-rules}
\end{figure*}

The program-logic rules for the probabilistic fragment of \thelang and presampling tapes are shown in \Cref{fig:probabilistic-rules}.
These rules reflect the operational semantics of \thelang.
For situations where we only want to progress the left program's random sampling without coupling, the rule \ruleref{wp-rand-l} can be used.
The rule \ruleref{wp-alloc-tape-l} allocates a fresh tape and returns its label.
The rules for sampling from a tape $\lbl$ depend on the contents of the tape: if the tape is \emph{not} empty, we pop and return the first value (\ruleref{wp-rand-tape-l}).
If the tape is empty, we sample an arbitrary integer from $0$ to $N$ (\ruleref{wp-rand-tape-empty-l}), just as for $\Rand~\tapebound$ without a tape annotation.

We emphasize that all of the rules shown so far are also found in the relational logic of Clutch~\cite{clutch}, \emph{except} that the right-hand-side rules in Clutch require that the left-hand-side program is \emph{not} a value.
This side-condition is a limitation of the model of Clutch that we eliminate in  \theaplog.
(This seemingly small improvement required \emph{significant} changes to the model, which we detail in \cref{sec:model}.)

\begin{figure*}[th!]
  \centering
  \begin{mathpar}
    \infrule{wp-rand-l}
    {
      \All n \leq \tapebound . \wpref{n}{\expr}{\Phi}
    }
    {\wpref{\Rand\tapebound}{\expr}{\Phi}}
    \and
    \infrule{wp-alloc-tape-l}
    { \All \lbl . \progtape{\lbl}{\tapebound}{\nil} \wand \wpref{\lbl}{\expr}{\Phi} }
    {\wpref{\AllocTape\tapebound}{\expr}{\Phi}}
    \and
    \infrule{wp-rand-tape-l}
    {
      \progtape{\lbl}{\tapebound}{n \cons \tape} \\
      \progtape{\lbl}{\tapebound}{\tape} \wand \wpref{n}{\expr}{\Phi}
    }
    {\wpref{\Rand\tapebound~\lbl}{\expr}{\Phi}}
    \and
    \infrule{wp-rand-tape-empty-l}
    {
      \progtape{\lbl}{\tapebound}{\nil} \\
      \All n \leq \tapebound . \progtape{\lbl}{\tapebound}{\nil} \wand \wpref{n}{\expr}{\Phi}
    }
    {\wpref{\Rand\tapebound~\lbl}{\expr}{\Phi}}
  \end{mathpar}
  \caption{A selection of program-logic rules of \theaplog for the probabilistic operations.}
  \label{fig:probabilistic-rules}
\end{figure*}

\paragraph{Approximate Coupling Rules}

The rules shown so far allow one to symbolically progress either the left- or right-hand side program of the weakest precondition assertion, independently of each other.
However to prove interesting relational properties, we need to progress the programs in a related manner using coupling rules, which we saw special cases of in \cref{sec:key-ideas}.

First, we have \ruleref{wp-couple-rand-rand-err-le} which relates sampling $\Rand \tapebound$ with $\Rand \tapeboundB$ where $\tapebound\leq\tapeboundB$.
Here $f: \mathbb{N}_{\leq \tapebound} \rightarrow \mathbb{N}_{\leq \tapeboundB}$ is an injective function ($\mathbb{N}_{\leq \tapebound}$ denotes the natural numbers $\leq \tapebound$) and by spending $\frac{\tapeboundB-\tapebound}{\tapeboundB+1}$ error credits, we may continue reasoning as if the return values are ``synchronized'' and related by $f$. This is also the rule we used for proving the switching lemma presented in \cref{sec:key-ideas}.
Note that in the special case where $N = M$, this generalizes the traditional coupling rule found in exact coupling logics (\eg{} Clutch~\cite{clutch}) where no error is incurred; the \ruleref{wp-couple-exact} rule in \cref{sec:key-ideas} is an example of this special case.

The \ruleref{wp-couple-rand-rand-err-ge} rule works almost identically, except that the inequality of the bound is reversed, \ie{}, $\tapebound \geq \tapeboundB$. (We mention that all the other rules we  present in this paper have symmetric versions, just as for \ruleref{wp-couple-rand-rand-err-le} and \ruleref{wp-couple-rand-rand-err-ge}. For the sake of brevity, we shall only present one direction of each pair of rules subsequently.)
\begin{mathpar}
  \inferH{wp-couple-rand-rand-err-le}
  {
    f: \mathbb{N}_{\leq \tapebound} \rightarrow \mathbb{N}_{\leq \tapeboundB}~\text{injection} \\
    \upto{\tfrac{\tapeboundB-\tapebound}{\tapeboundB+1}}\\
    \tapebound \leq \tapeboundB\\
    \All n \leq \tapebound .
    \wpref{n}{f(n)}{\Phi}
  }
  {\wpref{\Rand \tapebound}{\Rand \tapeboundB}{\Phi}}
  \and
  \inferH{wp-couple-rand-rand-err-ge}
  {
    f: \mathbb{N}_{\leq \tapeboundB} \rightarrow \mathbb{N}_{\leq \tapebound}~\text{injection} \\
    \upto{\tfrac{\tapebound-\tapeboundB}{\tapebound+1}}\\
    \tapebound \geq \tapeboundB\\
    \All n\leq\tapeboundB .
    \wpref{f(n)}{n}{\Phi}
  }
  {\wpref{\Rand \tapebound}{\Rand \tapeboundB}{\Phi}}
\end{mathpar}

As in Clutch, \theaplog also supports \emph{asynchronous} coupling.
For example, the \ruleref{wp-couple-tape-tape-err-ge} rule below is a variant of \ruleref{wp-couple-rand-rand-err-ge} where, instead of two program samplings, we couple two tape samplings.
\begin{mathpar}
  \inferH{wp-couple-tape-tape-err-ge}
  {
    f: \mathbb{N}_{\leq \tapeboundB} \rightarrow \mathbb{N}_{\leq \tapebound}~\text{injection} \\
    \tapebound \geq \tapeboundB\\
    \progtape{\lbl}{\tapebound}{\tape} \\
    \spectape{\lbl'}{\tapeboundB}{\vec{m}} \\
    \upto{\tfrac{\tapebound-\tapeboundB}{\tapebound+1}}\\
    \All n\leq\tapeboundB . \progtape{\lbl}{\tapebound}{\tape \cons f(n)} \sep \spectape{\lbl'}{\tapeboundB}{\vec{m} \cons n} \wand
    \wpref{\expr_1}{\expr_2}{\Phi}
  }
  {\wpref{\expr_1}{\expr_2}{\Phi}}

\end{mathpar}

\paragraph{Many-to-One and Fragmented Coupling Rules}
The coupling rules shown so far allow one to couple one sampling with another.
However, in certain cases, we may need to couple one sampling to \emph{zero} or \emph{multiple} possibly-non-adjacent samplings.
Consider the following two programs as an example: $\expr_1 \eqdef{} 2\cdot \Rand 1 + \Rand 1$ and $\expr_2 \eqdef{} \Rand 3$.
These programs are equivalent: $\expr_1$ samples two bits and returns the result interpreted in base $2$, while $\expr_2$ samples directly from the same distribution.
None of the coupling rules shown so far would allow us to relate these two programs.

Presampling tapes turn out to be a succinct and uniform solution to this problem, as smoothly enabled by the new and more flexible model of \theaplog. By reasoning about values stored in tapes, we can construct more intricate couplings that do not adhere to the one-to-one pattern exhibited in our previous rules.
This notion is captured by the following general rule:
\begin{mathpar}
  \inferH{wp-tape-tape-append}
         {\upto{\err} \\
           \progtape{\lbl}{\tapebound}{\tape} \\
           \spectape{\lbl'}{\tapeboundB}{\vec{m}} \\
           \ARcoupl{\unif(\tapebound+1)^p}{\unif(\tapeboundB+1)^q}{\err}{R} \\
           \All (v, w) \in R. \progtape{\lbl}{\tapebound}{\tape\dplus v} \sep \spectape{\lbl'}{\tapeboundB}{\vec{m}\dplus w} \wand \wpref{\expr_1}{\expr_2}{\Phi}
         }
         { \wpref{\expr_1}{\expr_2}{\Phi} }
\end{mathpar}
Here $\unif(x)^y$  refers to the uniform distribution of lists in $\textlog{List}(x,y)$, where $\textlog{List}(x,y)$ denotes lists of length $y$ containing integers not larger than $x$. Assume
we want to prove $\wpref{\expr_1}{\expr_2}{\Phi}$, and we are given $\upto{\err}$ error credits and tapes $\lbl$ and $\lbl'$ of bounds $\tapebound$ and $\tapeboundB$ on the left and right side of the refinement, respectively.
Then the rule says it suffices to (1) choose two lengths $p$ and $q$, and a relation $R$ over $\textlog{List}(N,p)$ and $\textlog{List}(M,q)$,
(2) prove an approximate coupling $\ARcoupl{\unif(\tapebound+1)^p}{\unif(\tapeboundB+1)^q}{\err}{R}$, and (3) show for all lists $(v, w) \in R$, the refinement weakest-precondition assertion holds after $v$ and $w$ are appended to some tapes $\lbl$ and $\lbl'$, respectively.

Intuitively, \ruleref{wp-tape-tape-append} is sound because appending lists sampled from the distribution $\unif(x)^y$ is an erasable action (see \cref{def:erasable}), meaning that it does not influence the probabilistic execution of any program. However, proving the asynchronous coupling $\ARcoupl{\mu_1}{\mu_2 }{\err_1}{R}$ for some arbitrary $\mu_1,\mu_2,$ and $R$ is generally not an easy task, so we provide various rules which are special instances of \ruleref{wp-tape-tape-append}.

First, we introduce the \ruleref{wp-many-to-one} rule, derived from \ruleref{wp-tape-tape-append}, that allows us to couple one sampling onto a tape with \emph{multiple} samplings onto another tape.
This allows us to handle the two programs $e_1$ and $e_2$ above that generate samples from $\{0, 1, 2, 3\}$.
\newcommand{\decoder}{\textlog{decoder}}
\begin{mathpar}
  \inferH{wp-many-to-one}
         { (N+1)^p = M+1 \\
             \progtape{\lbl}{\tapebound}{\tape} \\
             \spectape{\lbl'}{\tapeboundB}{\vec{m}} \\
           \All l. \length (l) = p \sep
           \progtape{\lbl}{\tapebound}{\tape\dplus l} \sep \spectape{\lbl'}{\tapeboundB}{\vec{m}\cons \decoder(N, l) } \wand
           \wpref{\expr_1}{\expr_2}{\Phi}
         }
         { \wpref{\expr_1}{\expr_2}{\Phi} }
\end{mathpar}
The meta-level function $\decoder$ takes as arguments an integer $N$ and a list of integers $l$ whose elements are smaller than or equal to $N$, and returns the integer represented by the list $l$ in base $N+1$.
For example $\decoder(1, [1,1, 0])$ returns the value $6$.
Intuitively, given that $(N+1)^p = M+1$, \ruleref{wp-many-to-one} couples $p$ samplings onto the tape $\lbl$ with a single sampling onto $\lbl'$, such that they are related by the $\decoder$ function.

In addition to many-to-one couplings, \theaplog also introduces a class of \emph{fragmented} coupling rules, which we briefly introduced in \cref{sec:key-ideas-rej-samp}. Fragmented coupling is a novel kind of coupling rule where the number of values inserted into the tapes are not uniform for all possible branches. Fragmented coupling rule are derived from a stronger notion of \ruleref{wp-tape-tape-append}, but the underlying principle is the same, in that the actions of inserting lists to various tapes are erasable and do not affect the probabilistic outcomes of program execution. The notion of fragmented couplings is captured by the following rule \ruleref{wp-fragmented-r-exp}:
\begin{mathpar}
  \inferH{wp-fragmented-r-exp}
  {
    f: \mathbb{N}_{\leq \tapebound} \rightarrow \mathbb{N}_{\leq \tapeboundB}~\text{injection} \\
    \tapebound<\tapeboundB \\
    \mhl{\upto{\err}} \\
      \progtape{\lbl}{\tapebound}{\tape} \\
      \spectape{\lbl'}{\tapeboundB}{\vec{m}} \\
    \All m \leq M .
    \spectape{\lbl'}{\tapeboundB}{\vec{m}\cons m} \sep
    \left(
    \text{if}~m \in \textlog{img}(f)
    \begin{array}[c]{l}
      \text{then}~\progtape{\lbl}{\tapebound}{\tape\cons f^{-1}(m)} \arcr
      \text{else}\;~\progtape{\lbl}{\tapebound}{\tape} \mhl{\sep \upto{\tfrac{\tapeboundB+1}{\tapeboundB-\tapebound}\cdot \err }}
    \end{array}
    \right)
    \wand
    \wpref{\expr_1}{\expr_2}{\Phi}
  }
  { \wpref{\expr_1}{\expr_2}{\Phi} }
\end{mathpar}
Ignore for now the highlighted assertions, the rule is also sound without them. The rule states that if we own tapes $\lbl$ and $\lbl'$ of type $N$ and $M$ where $N < M$, then for any injective function $f: \mathbb{N}_{\leq \tapebound} \rightarrow \mathbb{N}_{\leq \tapeboundB}$, we add the value $m$ to the $\lbl'$ tape and---if it exists---the pre-image $f^{-1}(m)$ to the $\lbl$ tape.
At first glance, this rule may seem arbitrary, but this conditional adding of a sample to one tape is crucial to reasoning about rejection samplers, as we saw in a simplified form in \cref{sec:key-ideas-rej-samp}.

Fragmented couplings can be generalized to approximate reasoning and expectation-preserving composition \cite{eris}, now also considering the highlighted assertions.
Namely, if we own $\err$ error credits we can distribute them uniformly across the branches that are not in the image of $f$.
That is, if the value is added to both tapes, no error is provided, and if a value is only added to the tape on the right-hand side, we pass the error amplified by a factor of $\tfrac{\tapeboundB+1}{\tapeboundB-\tapebound}$.

\paragraph{Error Amplification}
Recall that the standard recursion rule \ruleref{wp-rec} only works for recursive programs on the left-hand side of the refinement. To reason about recursive functions on the right-hand side of the refinement in \theaplog, one uses \emph{error amplification}, which was first introduced in the Eris logic~\cite{eris}.
\theaplog supports the following induction principle for error amplification.
\begin{mathpar}
  \infrule{err-amp}
  {
     0 < \err \\ 1 < k \\
    \forall \err'. (\upto{k \cdot \err'} \wand \prop ) \sep \upto{\err'} \vdash  \prop
  }
  {\upto{\err} \vdash\prop }
\end{mathpar}
The rule states that to prove $\prop$ given some positive error credits $\upto{\err}$, it suffices to prove $\prop$ given some arbitrary amount of error credits $\upto{\err'}$ and an inductive hypothesis for which we need for pay $\upto{k\cdot \err'}$ for some $k>1$. Intuitively, \ruleref{err-amp} is sound because given $k>1$, one can amplify any arbitrary positive error credit by $k$ repeatedly until the error reaches $1$, at which point we can derive $\FALSE$ by spending $1$ error credit. This induction principle encapsulates the kind of repeated amplification we alluded to at the end of \cref{sec:key-ideas-rej-samp}, avoiding the need to manually track how many rounds of amplification are needed.

We specialize the above rule to the following \ruleref{wp-err-amp} for reasoning about refinements:
\begin{mathpar}
  \inferH{wp-err-amp}
  {
     0 < \err \\ 1 < k \\
    \forall \err'. (\upto{k \cdot \err'} \wand \wpref{\expr}{\expr'}{\Phi} ) \sep \upto{\err'} \vdash  \wpref{\expr}{\expr'}{\Phi}
  }
  {\upto{\err} \vdash \wpref{\expr}{\expr'}{\Phi} }
\end{mathpar}

\subsection{Revisiting Rejection Samplers}
\label{sec:rej-samp}

Now that we have seen the rules of \theaplog, we return to the rejection sampler example from \cref{sec:key-ideas-rej-samp} and describe its proof in more detail.
Consider the two programs below, where $M<N$ (for now, it suffices to ignore the lines of code in \ghostcode{gray}), which reproduce the example from before.
{
  \begin{minipage}[t]{0.45\linewidth}
    \begin{align*}
      &\DumbLet \direct\ \vanon = \\
      &\tabw \ghostcode{\Let \lbl_{d} = \AllocTape \tapeboundB in} \\
      &\tabw\Rand \tapeboundB~\ghostcode{\iota_{d}}
    \end{align*}
  \end{minipage}
  \begin{minipage}[t]{0.48\linewidth}
    \begin{align*}
      &\DumbLet \RS\ \vanon = \\
      &\tabw\ghostcode{\Let \lbl_{r} = \AllocTape \tapebound in} \\
	    &\tabw(\Rec \vrs \vanon = \\
      &\tabw\tabw\Let \vx = \Rand \tapebound~\ghostcode{\lbl_{r}} in \\
	    &\tabw\tabw\If \vx \leq \tapeboundB then \vx \Else \vrs\ \TT)\ \TT
    \end{align*}
  \end{minipage}
  \vspace{1em}
}

On the left, $\direct$ is a simple program that samples directly from $\Rand M$.
On the right, $\RS$ is a rejection sampler. %
We aim to prove that they compute the same distribution.
To do so, we include extra code (in \ghostcode{gray}) to initialize and use presampling tapes. It is straightforward to use \theaplog to prove that the programs without tapes have the same execution distribution as their tape-annotated counterparts (which we omit for brevity).

By applying the \emph{error limiting adequacy result} (\cref{thm:lim-ade}), it is enough to assume we own an arbitrary and positive amount $\err$ of error credits, and prove the two following assertions:
\begin{align*}
	\err > 0 \sep \upto \err &\vdash \wpref {\direct\ \TT} {\RS\ \TT} {v,v' \ldotp v = v'}\, \\
	\err > 0 \sep \upto \err &\vdash \wpref {\RS\ \TT} {\direct\ \TT} {v,v' \ldotp v = v'}\,.
\end{align*}

We just show the first one, as the second one is mostly analogous (and in fact can be done without error credits by using the \rref{wp-rec} rule).
We begin by applying symbolic execution rules (\ruleref{wp-alloc-tape-l}) to allocate the tapes on both sides:
\begin{align*}
  \err > 0 \sep \upto \err \sep \progtape{\lbl_{d}}{\tapeboundB}{\nil} \sep \spectape{\lbl_{r}}{\tapebound}{\nil} \vdash \wpref {\Rand \tapeboundB~\lbl_d} {\vrs\ \TT} {v,v' \ldotp v = v'}\,.
\end{align*}

Although $\vrs$ is a recursive function, we cannot apply \ruleref{wp-rec} as it appears on the right-hand side. Instead, we leverage the error amplification proof technique and apply \ruleref{wp-err-amp}, with amplification factor $k \eqdef{} \frac{\tapebound+1}{\tapebound-\tapeboundB}$. Here $\Phi$ represents the inductive hypothesis we obtain:
\begin{align*}
	\Phi \sep \upto {\err'} \sep \progtape{\lbl_{d}}{\tapeboundB}{\nil} \sep \spectape{\lbl_{r}}{\tapebound}{\nil} &\vdash \wpref {\Rand \tapeboundB~\lbl_d} {\vrs\ \TT} {v,v' \ldotp v = v'} \\
	\text{where}~\Phi \eqdef{} \upto{k\cdot\err'} \sep \progtape{\lbl_{d}}{\tapeboundB}{\nil} \sep \spectape{\lbl_{r}}{\tapebound}{\nil} &\wand \wpref {\Rand \tapeboundB~\lbl_d} {\vrs\ \TT} {v,v' \ldotp v = v'}
  \end{align*}

We now continue by applying \ruleref{wp-fragmented-r-exp}, choosing $f \eqdef{} \lambda x. x$. This consumes our error credit $\upto{\err'}$ and distributes it unevenly across the branches depending on the sampling result. We then proceed with a case split.
In our first case, both tapes presample the same $v\leq M$:
\[\Phi \sep   \progtape{\lbl_{d}}{\tapeboundB}{[v]} \sep \spectape{\lbl_r}{\tapebound}{[v]} \vdash \wpref{\Rand \tapeboundB~ \lbl_d} {\vrs\ \TT}  {v, v' \ldotp v = v' } \]
Then, by taking primitive steps we return the same $v$ on both sides.

In our second case, we only push a value $v>M$ into the right-hand side tape $\lbl$ and we additionally have $\upto {k\cdot \err'}$ error credits:
\begin{align*}
  \Phi \sep \upto{k\cdot\err'}\sep \progtape{\lbl_d}{\tapeboundB}{\nil} \sep \spectape{\lbl_r}{\tapebound}{[v]} \vdash\ \wpref{\Rand \tapeboundB ~ \lbl_d} {\vrs\ \TT}  {v, v' \ldotp v = v' }
\end{align*}

We can now take steps only on the right-hand side. The sampling instruction will read $v$ from the tape, and the conditional will evaluate to the $\langkw{else}$ branch, which will leave us to prove:
\begin{align*}
  \Phi \sep \upto{k\cdot\err'}\sep \progtape{\lbl_d}{\tapeboundB}{\nil} \sep \spectape{\lbl_r}{\tapebound}{\nil} \vdash\ \wpref{\Rand \tapeboundB~\lbl_d} {\vrs\ \TT}  {v, v' \ldotp v = v' }
\end{align*}

Notice now we have amplified the error by a factor of exactly $k$, and hence we can directly apply $\Phi$, our hypothesis we obtained from \ruleref{wp-err-amp} to conclude the overall proof.

%% file: logrel.tex
\section{Logical Refinement}\label{sec:logrel}

It is often hard to reason directly about contextual equivalence, due to the
quantification over contexts. As in previous work~\cite{clutch} we define a
\emph{logical refinement relation} to help us reason about contextual
refinement. Like contextual refinement, logical refinement is a typed relation:
it ranges over pairs of expressions $e_1$, $e_2$ and types $\type$ such that
$e_1$ and $e_2$ have type $\type$. As we will show later, logical refinement
implies contextual refinement. However, logical refinement (as opposed to
contextual refinement) is defined in terms of the relational logic, 
and thus we can reason about it using the inference rules presented in previous
sections:
\begin{align*}
  \refines{\Delta}%
  {\expr_{1}}{\expr_{2}}{\type} \quad\eqdef{}\quad
  \All \varepsilon > 0 .
  &\upto{\err} \wand
  \wpref
    {\expr_{1}}
    {\expr_{2}}
    {\val_{1}, \val_{2} \ldotp \Exists \err' > 0 . %
	\upto{\err'} \sep \semInterp{\type}{\Delta}{\val_{1}, \val_{2}} } 
\end{align*}
Here, $\llbracket\type\rrbracket_{\Delta}$ denotes the semantic interpretation of type
$\type$, and $\Delta$ assigns a semantic interpretation to type variables in the
context. Intuitively speaking $e_1$ logically refines $e_2$ at type $\type$ if we can
couple their executions so that they return values related at the
semantic interpretation of $\type$. The key novelty with respect to prior work is the
quantification over error credits: we get to assume ownership of a positive
amount of error credits in our proof, as long as we ensure that at the end of
it we still have a positive amount left.
The quantification over error credits allows us to assume ownership of a
non-zero amount of error credits whenever reasoning about logical refinement,
in particular we can prove the following rule:
\vsquish{2pt}
  \begin{mathpar}
    \infrule[Right]{log-get-err}
           {
	     \forall \err > 0. \upto{\err} \wand \refines{\Delta}
             {e}{e'}{\type}
           }
	  { \refines{\Delta}
            {e}{e'}{\type} }
  \end{mathpar}
Since \rref{log-get-err} always allows us to obtain a positive amount of error credits, we can
internalize a closed rule for error induction for proving logical
refinements that assumes no previous ownership of error credits. This can be
seen as lifting the \ruleref{wp-err-amp} rule to the logical refinement level:
  \begin{mathpar}
    \infrule[Right]{log-ind-err}
           {
		   1 < k \\ \All \err . (\upto{k \cdot \err} \wand \refines{\Delta}{e}{e'}{\type} ) \wand \upto{\err} \wand  \refines{\Delta}{e}{e'}{\type}
           }
	  { \refines{\Delta}{e}{e'}{\type} }
  \end{mathpar}

The semantic interpretation $\semInterp{\type}{\Delta}{\val_{1},\val_{2}}$ of a type $\type$ relates
values (which do not need themselves to be syntactically well-typed) that behave \emph{as if} they were equivalent
values of type $\type$. This definition is mostly standard, and is defined as usual by induction on $\type$ and in mutual recursion with the refinement
relation, see Appendix A of~\cite{clutch-arxiv}, as well as~\cite{iris-logrel-journal} for a general account.
Semantic interpretation of a typing context
$\semInterp{\Gamma}{\Delta}{\gamma_1, \gamma_2}$ relates two substitutions $\gamma_1,\gamma_2$ whenever
for all $x \in {\sf dom}~\Gamma$, $\semInterp{\Gamma(x)}{\Delta}{\gamma_1(x), \gamma_2(x)}$.
Logical refinement can then be extended to open terms as usual:
\[ \refines{\Delta \mid \Gamma}{\expr_{1}}{\expr_{2}}{\type} \eqdef{} \forall \gamma_1, \gamma_2. \semInterp{\Gamma}{\Delta}{\gamma_1,\gamma_2} \wand \refines{\Delta}{\expr_{1}\gamma_1}{\expr_{2}\gamma_2}{\type} \]

We can prove a compatibility result, which intuitively states that all typing rules preserve the relation.
For example, in the case of function application we have:
\begin{equation*}
  \infrule[Right]{compat-app}{
	\refines{\Delta}{f_{1}}{f_{2}}{\type\to\typeB} \\
	\refines{\Delta}{\expr_{1}}{\expr_{2}}{\type}}
  {\refines{\Delta}{f_1~\expr_{1}}{f_2~\expr_{2}}{\typeB}}
\end{equation*}
The compatibility results can be combined into the fundamental lemma of the logical relation in the usual way, \ie, by induction on the typing derivation.
\begin{lemma}\label{lem:fundamental}
  If\; \(\Gamma \vdash \expr : \tau\)\; then\; \(\refines \Gamma \expr \expr \tau\)\,.
\end{lemma}

This logical refinement is a strict extension of the one in Clutch~\cite{clutch}: we can still prove the same
refinements (by not using the error credits at all), but we can also prove \emph{new} refinements
that were not provable in \loccit, in particular refinements involving recursive programs on the right.
The refinement relation is still sound with respect to contextual equivalence, as stated below:
\begin{theorem}[Soundness]
  Let $\Xi$ be a type variable context, and $\Delta$ a context assigning a relational interpretation to all type variables in $\Xi$.
  If $\refines{\Delta\mid\Gamma}{e_1}{e_2}{\type}{}$ then $\ctxrefines{\Xi\mid\Gamma}{e_1}{e_2}{\type}{}$.
\end{theorem}

The proof follows from compatibility of the logical refinement and the adequacy theorem of our relational logic.
In particular, we use \cref{thm:lim-ade} to erase the quantification over positive errors.

\begin{example}
	Logical refinement can be used to prove contextual equivalence of the two samplers considered in \cref{sec:rej-samp}. The statements we have to prove are below:
\begin{align*}
	\refines{\empty}{\direct}{\RS}{\tunit \to \tnat}, \qquad\qquad
	\refines{\empty}{\RS}{\direct}{\tunit \to \tnat}
\end{align*}
	Note that, as opposed to the proof in \cref{sec:rej-samp}, there is no need to assume ownership of a positive amount of error credits, and we can
	have a closed proof at the level of the logical refinement.
	
	Using similar ideas, \appref{app:vnd} presents a proof of contextual equivalence between a direct sampler over $\Rand~5$ (i.e., a die) and a rejection sampler that simulates it
	with 3 coin flips, encodes the result as a number from $0$ to $7$ and returns it if it is $5$ or less, retrying otherwise. In particular, this example uses our
	many-to-one coupling rules (\rref{wp-many-to-one}).
\end{example}

%% file: case-studies.tex
\section{Case Studies}
\label{sec:case-studies}

In this section we give an overview of several complex examples that we have verified using \theaplog. Complete details about each example can be found in the accompanying \rocq development.

\newcommand{\dsim}{\logv{dsim}}
\newcommand{\drej}{\logv{drej}}
\newcommand{\droll}{\logv{droll}}

\subsection{The PRP/PRF Switching Lemma, Revisited}
\label{sec:switching-lemma}

In \cref{sec:wsl}, we sketched a ``weak Switching Lemma'', where we considered one particular adversary.
We are now ready to prove the ``full'' version of the switching lemma for arbitrary adversary \(\adv\).
Of course, \emph{some} restrictions on the adversary are still required. First, the adversary must treat the RP/RF as a black box, and not directly access the underlying map that stores the function's values.
Second, we must ensure that the RP/RF can only be queried up to $Q$ times.
To enforce the first requirement, we require that \(\adv\) is a well-typed \thelang program.
For the second requirement, we wrap the RP/RF with the higher-order function \(\vbounded\), which uses local state that tracks how many queries have been performed:
\begin{align*}
  &\DumbLet \vbounded\ (Q : \tint)\ (\vf : \upalpha \ra \upbeta) : \upalpha \ra \upbeta\ \toption = \\
  &\quad\Let \vcounter = \Alloc 0 in \\
  &\quad\Fun \vx . \If (\deref \vcounter < Q)\ then\ \vincr\ \vcounter \Seq \Some\ (\vf\ \vx)\ \Else \None
\end{align*}
Our goal is then to prove the following logical refinement (as well as the other direction, the proof of which is similar and omitted) for any \(\adv\) of type \((\tint \ra \tint\ \toption) \ra \tbool\),
\begin{equation}\label{eq:sl-lr}
  \upto \err \wand\
  \refines {}
  {\adv\ (\vbounded\ Q\, (\erp\ \tapebound))}
  {\adv\ (\vbounded\ Q\, (\erf\ \tapebound))} \tbool
\end{equation}
where \(\err = {\frac{Q(Q-1)}{2\tapebound}}\). By unfolding the definition of the logical relation and applying the adequacy theorem of \theaplog to \eqref{eq:sl-lr}, these refinements imply \cref{lem:switching}.

In order to reason about the unknown program \(\adv\), we leverage the logical relation.
Specifically, from the assumption that \(\adv\) is well typed together with the Fundamental Lemma (\cref{lem:fundamental}), we derive an \theaplog specification for \(\adv\) of the form $\refines{}\adv\adv{((\tint\ra\tint\ \toption)\ra\tbool)}\label{eq:sl-lr-adv}$.
Using this and the logical relation's compatibility rule for function application leaves us to prove:
\begin{align}
  \upto \err \wand\ &\refines{}{(\vbounded\ Q\, (\erp\ \tapebound))}{(\vbounded\ Q\, (\erf\ \tapebound))}{\tint\ra\tint\ \toption} \label{eq:sl-lr-q}
\end{align}

In other words, after using the logical relation, the goal that remains makes no reference to the unknown code for the adversary, $\adv$.
From here on, the proof is very similar to that of the weak PRP/PRF Switching Lemma.
We symbolically evaluate both programs, which results in the allocation of the list of unused values in \(\erp\), the finite maps that both \(\erp\) and \(\erf\) use, and a counter for each program. Let \(\rp_Q\) denote the function returned by \(\vbounded\ Q\ (\erp\ \tapebound)\), and likewise \(\rf_Q\) for \(\erf\).
The key difference with the proof in \cref{sec:wsl} is that, instead of proving the \(\err\)-equivalence of two \(\langkw{for}\) loops by induction on \(Q\), we prove the \(\err\)-equivalence of two \emph{functions} by establishing an invariant that holds before and after all calls to  \(\rp_Q\) and \(\rf_Q\).
This invariant we need states that both counters point to the same value \(i\), that the maps \(\vm\) and \(\vm'\) are equal, and that the current error budget is \(\err_i = \sum_{k=i}^{Q-1}\frac k \tapebound\) when the counter is \(0 \leq i \le Q\). Furthermore, the list of unused values has length (at least) \(\tapebound-i\).
Since \(\upto{\err_i}\) is an ordinary proposition in \theaplog, and since \theaplog is an impredicative higher-order logic, we can simply ``store'' error credits in the invariant.

The semantic interpretation of function types requires us to prove that both functions in \eqref{eq:sl-lr-q} map related arguments to related results. In other words, we can show the refinement by applying both functions to the same integer \(n\).
Just as in \cref{sec:wsl}, we can then argue that with probability at most\footnote{This ``worst case'' occurs if all previous calls to \(\erp\) were made with different arguments and \(\listlen\ \vunused = \tapebound - i\).} \(\frac {\tapebound-(\tapebound-i)}{\tapebound} = \frac{i}{\tapebound}\) a collision occurs. By \rref{wp-couple-rand-rand-err-le}, we can force both functions to sample the same value by paying \(\upto{\frac i \tapebound}\). This is exactly the first element in the sum \(\err_i\). The functions thus return the same results, as required.
It remains to re-establish the function invariant. Since the counters have been incremented to \(i+1\), we only need to give back \(\err_{i+1}\) credits, which is exactly what is left of \(\upto{\err_i}\) after splitting off \(\upto{\frac i \tapebound}\).

In conclusion, thanks to the logical relation, proving the full Switching Lemma is as simple as proving the weak version.

\subsection{IND\$-CPA Security of Symmetric Encryption}
\label{sec:cpa}

A key notion of security for a symmetric (\ie, private key) encryption scheme is ``indistinguishability from random under chosen-plaintext attacks'' (IND\$-CPA, a.k.a.\ CPA\$). The IND\$-CPA ``advantage'' of an adversary \(\adv\) against an encryption scheme \((\vkeygen,\venc,\vdec)\) is defined as the probability that \(\adv\) is able to distinguish a ciphertext \(\vc\) corresponding to a plain-text message \(\vmsg\) from a randomly chosen ciphertext \(\vc'\), even if \(\adv\) can choose \(\vmsg\) (see, \eg, \cite[Def.~7.2]{Rosulek:Joy:2021}).
For an adversary that can make only up to $Q$-queries to the encryption oracle, the advantage is thus equal to
\begin{equation}
  \label{eq:cpa}
  \abs{\pr{\adv(\vbounded\ Q\,(\venc\, (\vkeygen\,\TT)))}
    - \pr{\adv(\vbounded\ Q\,(\vrandcipher))}}
\end{equation}
where \(\vrandcipher = \Fun \vmsg . (\Rand \tapebound, \Rand \tapebound)\) produces random ciphertexts.

It is well-known that a \emph{deterministic} encryption scheme cannot achieve IND\$-CPA security \cite{Katz:Introduction:2021,Rosulek:Joy:2021}.
A standard solution to obtain a IND\$-CPA secure scheme is to randomize the encryption function. We exemplify this idea by proving a bound on the IND\$-CPA advantage  for the following textbook construction \cite[Def.~7.4]{Rosulek:Joy:2021}\cite[Def.~3.28]{Katz:Introduction:2021}  of a symmetric scheme from a random function:
\begin{equation*}
  \begin{tabular}[t]{>{$}l<{$} @{ } >{$}c<{$} @{\; } >{$}l<{$} }
    \DumbLet \venc\ \prf\ \vkey\ \vmsg
    &\,=& \Let \vr = \Rand \tapebound in \\
    && \Let \vpad = \prf\ \vkey\ \vr in \\
    && \Let \vc = \vxor\ \vmsg\ \vpad in \\
    && (\vr, \vc)
  \end{tabular}
  \begin{tabular}[t]{>{$}l<{$} @{ } >{$}c<{$} @{\; } >{$}l<{$} }
    \DumbLet \vkeygen\ \TT &=& \Rand N \\
    \DumbLet \vdec\ \prf \ \vkey\ (\vr, \vc) &=
                           & \Let \vpad = \prf\ \vkey\ \vr in \\
                         && \Let \vmsg = \vxor\ \vc\ \vpad in \\
                         && \vmsg
  \end{tabular}
\end{equation*}
In particular, we prove the following refinement (and its converse) in \theaplog,
\begin{equation}\label{eq:cpa-lr}
  \upto{Q^2/(2\tapebound)} \wand\;
  \refines {}
  {\adv\ (\vbounded\ Q\ \venc_\rf)}
  {\adv\ (\vbounded\ Q\ \vrandcipher)}
  {\tbool}
\end{equation}
where \(\venc_\rf = \venc\ (\Let \rf = \erf\, \tapebound in \Fun \vkey . \rf)\).
Intuitively, the scheme is secure because \(\prf\) produces random-looking outputs. So long as \(\vpad = \prf\ \vkey\ \vr\) never repeats throughout the \(Q\) calls to \(\venc_\rf\), the \(\vxor\ \vmsg\ \vpad\) acts as a one-time pad, and the ciphertexts look random.

The proof of \eqref{eq:cpa-lr} thus hinges on the fact that the randomly sampled value \(\vr\) never repeats, since this ensures that \(\erf\) samples a new value for \(\vpad\).
Formally, we argue that after initialization of \(\erf\) and \(\vbounded\), the encryption oracle \(\venc_\rf\) and the random cipher oracle satisfy the following invariant, which ties the amount of error credits left to the counter value \(i\).
\begin{equation}
  \label{eq:cpa-inv}
  \Exists i, m .  \upto{(Q^2-i^2)/(2\tapebound)}
  \sep \vcounter \pointsto i
  \sep \vcounter' \spointsto i
  \sep \isprf m
  \sep \card{\dom\ m} = i
\end{equation}
We use the approximate coupling rule to ``pay off'' the risk of a repeated use of \(\vr\) at the \(i\)-th oracle call.
The exact source of errors is different here from the Switching Lemma, since we have to argue that the randomly sampled arguments to \(\rf\) do not repeat, whereas in the Switching Lemma, we are concerned with collisions that get sampled if fresh arguments are fed to \(\rf\) and \(\rp\).
The exact rule that allows us to do this is given below, where the list \(l\) is instantiated with the domain of the map \(m\) which is tracked by \(\rf\) in the invariant  \eqref{eq:cpa-inv}.
\begin{mathpar}
  \infrule*{}
  {
    \upto{\tfrac{\length(l)}{\tapebound+1}}\\
    \All n\leq\tapebound . n\notin l \wand
    \wpref{n}{n}{\Phi}
  }
  {\wpref{\Rand \tapebound}{\Rand \tapebound}{\Phi}}
\end{mathpar}
Since the invariant is preserved throughout, the output of \(\venc_\rf\) is thus \(\frac{Q^2}{2\tapebound}\)-equivalent to that of \(\vrandcipher\) in \(Q\) oracle calls.

\subsection{Sampling from B+ Trees}
\label{sec:sampling-b-trees}
\newcommand{\maxchildnum}{M}
\newcommand{\totalleavesnum}{\logv{total\_leaves\_num}}
\newcommand{\naiveprog}{\langv{naive\_sample}}
\newcommand{\optimizedprog}{\langv{optimized\_sample}}
\newcommand{\optimizedsearch}{\langv{draw}}
\newcommand{\createtree}{\langv{create\_tree}}
\newcommand{\createrankedtree}{\langv{create\_ranked\_tree}}
\newcommand{\naiveannotatedprog}{\langv{naive\_annotated\_sample}}
\newcommand{\optimizedannotatedprog}{\langv{optimized\_annotated\_sample}}
\newcommand{\intermediateprog}{\langv{intermediate\_sample}}
\newcommand{\getdepth}{\logv{get\_depth}}
\newcommand{\depth}{\logv{d}}
\newcommand{\tree}{\logv{tree}}
\newcommand{\treetype}{\tau}
\newcommand{\naivepack}{\langv{naive}}
\newcommand{\optpack}{\langv{opt}}
\newcommand{\intty}{\textsf{Int}}

\newcommand{\inittree}{\langv{init\_tree}}
\newcommand{\inserttree}{\langv{insert\_tree}}
\newcommand{\buildranked}{\langv{build\_ranked}}

In this case study, we show the correctness of a rejection sampling scheme developed by \citet{b+_tree_sample} for drawing a random sample from a B+ tree.
This case study demonstrates how \theaplog is able to handle complex mutable state and establish equivalences that rely on type abstraction.
A B+ tree~\citep{b+_tree} is a height-balanced tree data structure that is widely used for storing data in filesystems and databases.
Unlike a binary search tree, a B+ tree's internal nodes may have more than 2 children, up to some maximum $\maxchildnum$.

If a B+ tree's nodes include additional \emph{ranking} information recording how many leaves are descendants of each node, then it is straight-forward to draw a random element.
For a tree with $N$ total elements, draw a random number uniformly from $\{0, \dots, N-1\}$ and then use the ranks to find the $i$-th element in the tree.
However, maintaining the ranks has overhead.

\citet{b+_tree_sample} developed a rejection sampling algorithm for sampling from a \emph{non}-ranked B+ tree.
Starting from the root, the algorithm recursively descends down the tree.
At each non-leaf node, it samples a random number $i$ uniformly from $\{0, 1, \dots, \maxchildnum-1\}$.
If the node has an $i$-th child, the algorithm recurses on it.
If the node does not have an $i$-th child, it aborts early by returning to the root and restarting.
Once the algorithm reaches a leaf, it returns it as the selected sample.

We have implemented the sampling algorithm for ranked B+ trees and the rejection sampler for non-ranked trees as two functions called $\naiveprog$ and $\optimizedprog$, respectively (see \appref{app:sampling-b-trees} for the full code).
Our main result for this case study shows that these two functions are equivalent.
Of course, they are only equivalent when they operate over well-formed trees, so we state this result as a contextual equivalence about two different implementations of an abstract tree data type.
To do so, we first define the following additional functions: $\inittree$, which takes an integer and returns a B+ tree containing that integer, $\inserttree$, which inserts an integer into a tree, and $\buildranked$, which takes a (non-ranked) B+ tree and returns a ranked B+ tree with the same entries and shape.
Then the following two packed tuples bundle the B+ tree operations:
\begin{align*}
\optpack &\eqdef \Pack (\inittree, \inserttree, \optimizedprog) \\
\naivepack &\eqdef \Pack (\inittree, \inserttree, \lambda t. \naiveprog\ (\buildranked\ t))
\end{align*}
where the sampling routine in $\naivepack$ takes a tree $t$, builds the ranked version of the tree, and then uses the na\"ive routine.\footnote{Building a ranked tree each time is inefficient, but $\naivepack$ serves as a specification for $\optpack$, so its efficiency is not relevant.}
With these preliminaries in place, our main result can be stated as $\ctxeq{}{\optpack}{\naivepack}{\treetype}{}$, where $\treetype \eqdef \Exists \tau. (\intty \rightarrow \tau) \times (\tau \times \intty \rightarrow \unitty) \times (\tau \rightarrow \intty)$.

This proof is described in more detail in \appref{app:sampling-b-trees}.
At a high level, it has two components.
First, there is non-probabilistic reasoning showing that the various routines traverse and modify the trees correctly.
This makes up the bulk of the proof and consists of traditional separation-logic style reasoning about trees.
For this part, \theaplog{}'s support for the rich reasoning principles developed in earlier separation logics is essential.
The second component is the actual probabilistic reasoning using couplings.
Here, the coupling reasoning in this proof is quite similar to the arguments we have already seen using fragmented and many-to-one couplings in simpler rejection samplers.

\newcommand{\expsymb}{\mathrel{\textasciicircum}}

%% file: model.tex
\section{Semantic Model and Soundness}
\label{sec:model}
The soundness of \theaplog{} is justified by defining a semantic model of $\wpref{\expr}{\expr'}{\pred}$ in the Iris base logic \cite{irisjournal}.
The base logic is a higher-order separation logic that lacks any connectives for reasoning about programs.
In this section, we define the semantic model of $\wprenameref$ and discuss how it implies the existence of an approximate coupling of the execution of the two programs.

The model of $\wprenameref$ is inspired by the model of the (non-approximate) coupling logic Clutch \cite{clutch}.
We emphasize the following technical novelties and improvements over Clutch's model:
\begin{enumerate}
\item The approach is generalized to approximate couplings and expectation-preserving composition by incorporating error credits \cite{eris} in the relational setting.
\item The model allows for coupling rules and right-hand side rules to be applied when the left-hand side is a value (a limitation of the general structure of the Clutch model).
\item The model introduces two new \emph{coupling precondition} connectives and a notion of \emph{erasability} that captures the essence of why asynchronous couplings \cite{clutch} are sound.
\end{enumerate}

\subsection{Model}
The semantic model of $\wprenameref$ is defined using two unary connectives: a weakest precondition $\wpre{\expr}{\Phi}$ and a separation-logic resource $\spec(\expr')$ that tracks the right-hand-side specification program, as in prior work on refinement reasoning in separation logic~\citep{caresl,iris-logrel-journal, clutch}.
The $\wprenameref$ is defined as
\[
  \wpref{\expr}{\expr'}{\pred} \eqdef{}
  \All \lctx . \spec(\fillctx \lctx[\expr']) \wand
  \wpre \expr {\val \ldotp \Exists \val' . \spec(\fillctx \lctx[\val']) \sep \pred(\val, \val') }.
\]
By quantifying over evaluation contexts $\lctx$, we close the definition under evaluation contexts on the right-hand side; for the left-hand side this is not needed as the weakest precondition already satisfies the bind rule.
The main challenge of defining the relational connective is thus to define the model of the unary weakest precondition in a suitable way.

In isolation, the weakest precondition $\wpre{\expr}{\pred}$ encodes partial correctness: intuitively it means that the execution of $\expr$ is \emph{safe} (\ie, the probability of crashing is zero) and for every possible return value $\val$, the postcondition $\pred(\val)$ holds.
Internally, however, in order to do approximate (relational) reasoning, the weakest precondition pairs up of the probability distribution of individual steps of the program with the probability distribution of individual steps of \emph{some} other program, in such a way that there exists an approximate coupling among them.
Through separation-logic machinery, we tie this ``other'' program to the program tracked by the $\spec(\expr')$ resource, and the approximation error to error credits $\upto{\err}$.
The weakest precondition itself satisfies all the usual program logic rules that one would expect and we refer to \cref{fig:wp-rules} for an overview.

\paragraph{Weakest Precondition}

The definition of the weakest precondition is shown below.
As done throughout this paper, we ignore the general connectives that are used for manipulating  Iris-style ghost resources and invariants, \ie{}, the \emph{update modality} \(\ghostcode{\pvs}\) and \emph{invariant masks} (\(\ghostcode{\mask}\), \(\ghostcode{\emptyset}\)) as found in Iris \cite{irisjournal}, which are orthogonal to the core challenges that we address.
Our use is standard and the weakest precondition can be understood by omitting the grayed out parts.
The definition looks as follows.
\begin{align*}
  \wpre{\expr_{1}}[\ghostcode{\mask}]{\pred} \eqdef{}
  & \All \state_{1}, \cfg_{1}', \err_{1} .
    \stateinterp(\state_{1}, \cfg_{1}', \err_{1}) \wand \\
  & \quad \ghostcode{\pvs[\mask][\emptyset]} \specCoupll{\state_{1}}{\cfg_{1}'}{\err_{1}}\spac\{ \state_{2}, \cfg_{2}', \err_{2} \ldotp \\
  & \qquad \big(\expr_{1} \in \Val \sep
    \ghostcode{\pvs[\emptyset][\mask]} \stateinterp(\state_{2}, \cfg_{2}', \err_{2}) \sep \pred(\expr_{1})\big) \lor{} \\
  & \qquad
    \big( \expr_{1} \not\in \Val \sep
    \progCoupll{(\expr_{1}, \state_{2})}{\cfg_{2}'}{\err_{2}}
    \spac\{ \expr_{2}, \state_{3}, \cfg_{3}', \err_{3} \ldotp  \\
  & \quad\qquad \later \specCoupll{\state_{3}}{\cfg_{3}'}{\err_{3}}
    \spac\{ \state_{4}, \cfg_{4}', \err_{4} \ldotp
    \ghostcode{\pvs[\emptyset][\mask]} \stateinterp(\state_{4}, \cfg_{4}', \err_{4}) \sep \wpre{\expr_{2}}[\ghostcode{\mask}]{\pred} \} \} \big) \}
\end{align*}
The connective is defined as a \emph{guarded fixed point} of the equation above, as is custom in many program logics for partial correctness.
The fixed point exists because the recursive occurrence of the weakest precondition occurs under the later modality $\later$~\cite{iris2}.

The diagram below shows how the two programs are updated in a single unrolling of the weakest precondition. This will contain a single
execution step on the left-hand side, preceded and followed by a sequence of execution steps on the right-hand side. Along these steps,
the tapes can be updated with new samples, with errors threaded through to construct approximate couplings between steps:
\begin{center}
  \tikzset{every loop/.style={min distance=5mm,in=180,out=140,looseness=2}}
  \begin{tikzpicture}[node distance=3.75cm,on grid, auto]
    \node [inner sep=10pt] (A) {$(\expr_1,\state_1) \sim_{\err_1} \cfg_1'$};
	  \node [inner sep=10pt, right of=A] (B) {\;$(\expr_1,\state_2) \sim_{\err_2} \cfg_2'$\;};
    \node [inner sep=10pt, right of=B] (C) {$(\expr_2,\state_3) \sim_{\err_3} \cfg_3'$};
    \node [inner sep=10pt, right of=C] (D) {$(\expr_2,\state_4) \sim_{\err_4} \cfg_4'$};
    \path[->] (A) edge [decoration=snake] node[above=2mm]
    {\parbox{2.1cm} {\footnotesize \centering RHS $\exec_n$, \\ update tapes}} 
	node[pos=.5,below=2mm] {\footnotesize $\textlog{scpl}$} (B);
    \path[->] (B) edge [decoration=snake] node[above=2mm]
    {\parbox{2.8cm} {\footnotesize \centering LHS $\exec_1$, RHS $\exec_n$, \\ update tapes}} 
	node[pos=.5,below=2mm] {\footnotesize $\textlog{pcpl}$} (C);
    \path[->] (C) edge [decoration=snake] node[above=2mm]
    {\parbox{2.1cm} {\footnotesize \centering RHS $\exec_n$, \\ update tapes}} 
	node[pos=.5,below=2mm] {\footnotesize $\textlog{scpl}$} (D);
  \end{tikzpicture}
\end{center}

We now explain the definition in detail.
We first start by assuming ownership of a \emph{state interpretation} $\stateinterp(\state_{1}, \cfg_{1}', \err_{1})$.
The state interpretation predicate $ S : \State \to \Cfg \to [0,1] \to \iProp$ interprets the physical state of the program, the specification program, and the approximation error as resources in \theaplog{} which, \eg{}, gives meaning to the points-to connective $\progheap{\loc}{\val}$, the specification resource $\spec(\expr')$, and error credits $\upto{\err}$.
Our choice of resource algebras is standard (see \citet{clutch} and \citet{eris} for more details) and it is sufficient to know that they reflect the (partial) knowledge that the logical connectives represent.
For instance, for the heap points-to connective we have that $\stateinterp(\state_{1}, \cfg_{1}', \err_{1}) \sep \progheap{\loc}{\val} \vdash \state_{1}(\loc) = \val$; for the specification resource we have $\stateinterp(\state_{1}, (\expr_{1}', \state_{1}'), \err_{1}) \sep \spec(\expr') \vdash \expr_{1}' = \expr'$; and for error credits we have $\stateinterp(\state_{1}, \cfg_{1}', \err_{1}) \sep \upto{\err} \vdash \err_{1} \geq \err$.

Second, we have to prove a \emph{spec-coupling precondition} $\specCoupl{\state_1}{\cfg_{1}'}{\err_1}{\ldots}$.
We define the connective in \cref{fig:specCoupl}, but in essence it allows the \emph{right-hand-side} program to be progressed.
Intuitively, $\specCoupl{\state_1}{\cfg_{1}'}{\err_1}{\state_{2}, \cfg_{2}', \err_{2} \ldotp \prop}$ says that with error budget $\err_{1}$ there exists a (possibly empty) sequence of composable approximate couplings starting from state $\state_{1}$ and configuration $\cfg_{1}'$ that ends up in some state $\state_{2}$ and configuration $\cfg_{2}'$ with leftover error budget $\err_{2}$, such that the proposition $\prop$ holds.
By allowing the left-hand-side \emph{state} to be progressed with the right-hand-side \emph{configuration}, we permit certain \emph{asynchronous} coupling rules as discussed below in detail.

Next, if $\expr_{1}$ is a value we have to return the updated state interpretation and prove the postcondition $\pred(\expr_{1})$.
If $\expr_{1}$ is \emph{not} a value, we have to prove a \emph{program-coupling precondition} $\progCoupl{(\expr_{1}, \state_{2})}{\cfg_{2}'}{\err_{2}}{\ldots}$.
We define the connective formally below, but it allows the \emph{left-hand-side} program to take a single step and the
\emph{right-hand-side} program to take a finite number of steps.
Intuitively, $\progCoupl{\cfg_{1}}{\cfg_{1}'}{\err_1}{\cfg_{2}, \cfg_{2}', \err_{2} \ldotp \prop}$ says that with error budget $\err_{1}$ there exists an approximate coupling of a \emph{single} step of configuration $\cfg_{1}$ with a finite number of steps of configuration $\cfg_{1}'$ that ends up in configurations $\cfg_{2}$ and $\cfg_{2}'$ with leftover error budget $\err_{2}$, such that the proposition $\prop$ holds.

Finally, under a later modality (which signifies that a step of $\expr_{1}$ has been taken), we have to prove another spec-coupling precondition before returning the updated state interpretation and showing that $\wpre{\expr_{2}}{\pred}$ holds recursively.
The second occurrence of the spec-coupling precondition can mostly be ignored and is only required to validate the invariant opening rule which we omit.

\paragraph{Coupling Preconditions}
\begin{figure*}[t]
  \centering
  \begin{mathpar}
    \inferH{spec-coupl-err-1}
    {}
    { \specCoupl{\state}{\rho}{1}{\Phi} }
    \and
    \inferH{spec-coupl-ret}
    { \Phi(\state, \cfg', \err) }
    { \specCoupl{\state}{\cfg'}{\err}{\Phi} }
    \and
    \inferH{spec-coupl-exp}
    {
      \ARcoupl{\mu_1}{(\mu_1' \mbindi \Lam \state_2' . \pexec_n(\expr_1', \state_2'))}{\err_1}{R} \\
      \err_1 + \expect{\Err_2} \leq \err \\
      \erasable(\mu_1, \state_1) \\
      \erasable(\mu_1', \state_1') \\
      \All (\sigma_2, \cfg_2') \in R . \ghostcode{\pvs[\emptyset]} \specCoupl{\state_2}{\cfg_2'}{\Err_2(\cfg_2')}{\Phi}
    }
    { \specCoupl{\state_1}{(\expr_1', \state_1')}{\err}{\Phi} }
  \end{mathpar}
  \caption{Inductive definition of the specification-coupling precondition $\specCoupl{\state}{\rho}{\err}{\Phi}$.}
  \label{fig:specCoupl}
\end{figure*}

The spec-coupling precondition is defined inductively by the inference rules shown in \cref{fig:specCoupl}.
If the error budget is $1$ or if the postcondition holds for the input parameters, the spec-coupling precondition holds trivially (\ruleref{spec-coupl-err-1} and \ruleref{spec-coupl-ret}, respectively).
The last constructor (\ruleref{spec-coupl-exp}) is by far the most interesting: it allows us to incorporate approximate couplings and requires the existence of an $(\err_{1}, R)$-coupling of $\mu_{1}$ and $(\mu_1' \mbindi \Lam \state_2' . \pexec_n(\expr_1', \state_2'))$ for prover-chosen $\mu_{1}$, $\mu_{1}'$ and $n$.
Here $\pexec_{n}\colon \Conf \to \Distr{\Conf}$ denotes $n$ steps of partial execution, \ie{},
\begin{align*}
  \pexec_{n}(\expr, \state) \eqdef{}
  \begin{cases}
    \mret(\expr, \state)                                         & \text{if\;}\expr \in \Val \text{\; or\; } n = 0, \\
    \stepdistr(\expr, \state) \mbindi \pexec_{(n - 1)} & \text{otherwise.}
  \end{cases}
\end{align*}
and $\mu_{1}, \mu_{1}' \in \Distr{\State}$ are arbitrary \emph{erasable} distributions (\cf{}, \cref{def:erasable}) w.r.t. $\state_{1}$ and $\state_{1}'$.

Finally, to allow expectation-preserving composition (\cf{}, \cref{lem:arcoupl-exp-r}), the prover picks an error function $\Err_{2}$ such that $\err_1 + \expect{\Err_2} \leq \err$, where the expectation $\expect{\Err_2}$ is computed with respect to the distribution $(\mu_1' \mbindi \Lam \state_2' . \pexec_n(\expr_1', \state_2'))$.
Then for all $\sigma_2$ and $\cfg_{2}'$ in the support of the coupling, the spec-coupling precondition must hold recursively with the new error budget $\Err_{2}(\cfg_{2}')$.

The program-coupling precondition is defined in a similar style to the \ruleref{spec-coupl-exp} constructor, but the approximate coupling requires \emph{exactly} one step on the left-hand side as seen below.
\begin{mathpar}
  \infer
  { \ARcoupl{\stepdistr(\expr_1, \state_1)}{(\mu_1' \mbindi \Lam \state_2' . \pexec_{n}(\expr_1', \state_2'))}{\err_1}{R} \\
    \red(\expr_1, \state_1) \\
    \err_1 + \expect{\Err_2} \leq \err \\
    \erasable(\mu_1', \state_1') \\
    \All ((\expr_2, \sigma_2), (\expr_2', \sigma_2')) \in R . \ghostcode{\pvs[\emptyset]} \Phi(\expr_2, \state_2, \expr_2', \state_2', \Err_2(\expr_2, \state_2)) }
  {\progCoupl{(\expr_1, \state_1)}{(\expr_1', \state_1')}{\err}{\Phi}}
\end{mathpar}
The left-hand side program is also required to be reducible (to guarantee safety), and for all configurations in the support of the coupling, the postcondition must hold.
Note that the expectation $\expect{\Err_2}$ is taken with respect to the distribution $\stepdistr(\expr_1, \state_1)$.
This guarantees that every recursive unfolding of the weakest precondition corresponds to a single step of the left-hand-side program which is essential to validating the standard program logic rules found in \cref{fig:wp-rules}.

\begin{figure}[htb]
  \centering
  \begin{align*}
    \expr_{1} \purestep \expr_{2} \ast \wpre{\expr_{2}}{\pred} &\proves \wpre{\expr_{1}}{\pred} \\
    \All \loc . \progheap{\loc}{\val} \wand \pred(\loc) &\proves \wpre{\Alloc \val}{\pred} \\
    (\progheap{\loc}{\val} \wand \pred(\val))\sep {\progheap{\loc}{\val}} &\proves \wpre{\deref\loc}{\pred}  \\
    (\progheap{\loc}{\valB} \wand \pred(\TT))\sep {\progheap{\loc}{\val}} &\proves \wpre{\loc \gets \valB}{\pred}  \\
    \All n \leq \tapebound . \pred(n) &\proves \wpre{\Rand \tapebound}{\pred} \\
    \pred(\val) &\proves \wpre{\val}{\pred} \\
    \wpre{\expr}{\val .\, \wpre{\fillctx\lctx[\val]}{\pred}} &\proves \wpre{\fillctx\lctx[\expr]}{\pred}  \\
    (\All \val . \Psi(\val) \wand \pred(\val)) \sep \wpre{\expr}{\Psi} &\proves \wpre{\expr}{\pred}  \\
    \prop \sep \wpre{\expr}{\pred} &\proves \wpre{\expr}{v . \, \prop \sep \pred(v)} \\
    \All \lbl . \progtape{\lbl}{\tapebound}{\nil} \wand \pred(\lbl) & \proves \wpre{\AllocTape \tapebound}{\pred} \\
    (\All n \leq \tapebound . \progtape{\lbl}{\tapebound}{\nil} \wand \pred(n)) \sep \progtape{\lbl}{\tapebound}{\nil} &\proves \wpre{\Rand \tapebound~\lbl}{\pred} \\
    (\progtape{\lbl}{\tapebound}{\tape} \wand \pred(n)) \sep \progtape{\lbl}{\tapebound}{n \cons \tape} &\proves \wpre{\Rand \tapebound~\lbl}{\pred}
  \end{align*}
  \caption{Standard weakest-precondition rules.}
  \label{fig:wp-rules}
\end{figure}

The lemmas below illustrate how spec-coupling and program-coupling preconditions interact with the operational semantics to allow us to construct couplings for program executions.
First, we see the case of the spec-coupling precondition:

\begin{lemma}\label{lem:adeq-spec-coupl}
  Let \((\expr, \state_1)\) and \(\cfg_1'\) be configurations for the left-hand-side and right-hand-side programs, and let \(\varphi \subseteq \Val \times \Val\) be a relation on values.
  If, for some error \(\err_1 \in [0,1]\), 
  \[
    \specCoupl{\state_1}{\cfg'_1}{\err_1}{\state_2, \cfg'_2, \err_2 \ldotp \ARcoupl{\exec_{m}(\expr, \state_2)}{\exec(\cfg'_2)}{\err_2}{\varphi}} ,
  \]
  then there exists a \((\err_1, \varphi)\)-coupling $\ARcoupl{\exec_{m}(\expr, \state_1)}{\exec(\cfg'_1)}{\err_1}{\varphi}$.
\end{lemma}

The program-coupling precondition satisfies an analogous result, but notice the extra computation
step in the conclusion:

\begin{lemma}\label{lem:adeq-prog-coupl}
  Let \((\expr_1,\state_1)\) and \(\cfg_{1}'\) be configurations for the left-hand-side and right-hand-side programs where \(\expr_1 \not\in \Val\), and let \(\varphi \subseteq \Val \times \Val\) be a relation on values.
  If, for some error \(\err_1\),
  \[
    \progCoupl{(\expr_1,\state_1)}{\cfg_{1}'}{\err_1}{\expr_2, \state_2, \cfg'_2, \err_2 \ldotp \ARcoupl{\exec_{m}(\expr_2, \state_2)}{\exec(\cfg'_2)}{\err_2}{\varphi}} ,
  \]
  then there exists a \((\err_1, \varphi)\)-coupling $\ARcoupl{\exec_{m+1}(\expr_1, \state_1)}{\exec(\cfg_{1}')}{\err_1}{\varphi}$.
\end{lemma}

The proofs of these auxiliary lemmas rely on erasability as well as \cref{lem:arcoupl-exp-l,lem:arcoupl-exp-r} to construct the coupling of the executions.

\subsection{Soundness}
Soundness of \theaplog{} and the relational program logic follows from the adequacy theorem below.
\begin{theorem}[Adequacy]
  \begin{sloppypar}
  \ Let $\varphi \subseteq \Val \times \Val$ be a relation over values and let $0 \leq \err \leq 1$.
  If
  ${\spec(\expr') \sep \upto{\err} \vdash \wpre{\expr}{\val \ldotp \Exists \val'. \spec(\val') \ast \varphi(\val, \val')}}$
  then $\All \state, \state'. \ARcoupl{\exec(\expr, \state)}{\exec(\expr', \state')}{\err}{\varphi}$.
  \end{sloppypar}
\end{theorem}
The proof has a similar structure to the soundness theorem of Clutch \cite{clutch}. By continuity,
it suffices to show the following approximate coupling:
\[
\ARcoupl{\exec_{n}(\expr, \state)}{\exec(\expr', \state')}{\err}{\varphi}
\]
for all $\state$, $\state'$, and $n$.
The theorem then follows by induction on $n$.
The interesting case is the inductive step, when $n = m+1$.
After unfolding the definition of the weakest precondition, we can apply \cref{lem:adeq-spec-coupl} and \cref{lem:adeq-prog-coupl} to construct a coupling between $\exec_{m+1}(\expr, \state)$ and $\exec(\expr', \state')$.

%% file: work.tex
\section{Related Work}
\label{sec:work}

\paragraph{Probabilistic Couplings}
Relational reasoning about program via (exact) probabilistic couplings can be traced back to pRHL~\cite{pRHL,relational_via_prob_coupling} and was later extended to support approximate couplings in apRHL~\cite{apRHL,advanced_for_diff_privacy}, apRHL+~\cite{apRHL+}, and EpRHL~\cite{EpRHL}.
These approximate logics can be used to reason about a wide range of properties such as differential privacy and expected sensitivity, but they are limited to  reasoning about first-order programs.
\citet{HORHL} introduce HO-RHL, which use couplings to reason about adversarial computations in a higher-order setting.
HO-RHL, however, only allows synchronous couplings and only supports first-order global state and structural recursion.
Clutch~\citep{clutch} introduces asynchronous couplings in a higher-order setting with higher-order local state.
However, Clutch does not support approximate or fragmented couplings.

\paragraph{Approximate Reasoning}
Aside from relational approaches, approximate reasoning has also been used in the unary setting.
The unary logic aHL~\cite{aHL} is used to reason about accuracy properties of first-order randomized algorithms, where errors are tracked by a grading on Hoare triples.
Eris~\cite{eris} extends this to the higher-order setting and tracks error probability as a separation logic resource. 
In a slightly different line of work, expectation-based logics~\cite{WP-expected, prob-pred-trans, QSL,aert} can also be used to reason about approximate correctness of first-order imperative probabilistic programs. In particular, eRHL~\cite{eRHL:arxiv} supports reasoning about asynchronous samplings via \(\star\)-couplings.

\paragraph{Resource Reasoning with Credits}
Using sub-structural credits to track a program's resource consumption was pioneered in type systems for automated amortized resource analysis~(AARA)~\citep{static_prediction_heap}.
Subsequent research extends this approach to reason about expected cost bounds in probabilistic programs~\cite{bounded_expectations,raising_expectations,probabilistic_session_types}.
Inspired by AARA, \citet{resource-separation} introduced resource-tracking credits in separation logic to reason about amortized resource consumption.
A variant of this idea is implemented as time credits in Iris to reason about running time complexity of higher-order programs~\cite{time-credit-union-find,time-credit,time-credit-thunk} and expected running time in Tachis~\cite{tachis}.
Eris~\cite{eris} uses error credits to reason about error bounds of higher-order probabilistic programs, which \theaplog adapts to the relational setting.

\paragraph{Logical Relations and Probability}
\citet{LR_probability} developed a logical relations model of a higher-order probabilistic programming languages involving both state and discrete probabilistic choice to reason contextual equivalence.
The approach was then extended to support continuous probabilistic choice~\cite{CE_continuous, wand}, recursively nested queries~\cite{RARAR}, and non-determinism~\cite{LR_nondeterminism}.
The logical relation developed in Clutch~\cite{clutch} supports asynchronous couplings and is very similar to the model in our work.
Our logical relation, however, also supports proving contextual equivalences by means of approximation.
This is key to proving equivalences of rejection sampling programs which, to our knowledge, is out of scope for previous models based on logical relations.

Besides contextual equivalence, logical relations are used to reason about contextual distance between probabilistic programs~\cite{metric_reasoning, metric_reasoning_affine}.
Contextual distance can be seen as a generalization of contextual equivalence into a metric for analyzing distances between probabilistic programs.
Using error credits to reason about contextual distances is an interesting avenue for future work.

\paragraph{Separation Logic and Probability}

In addition to Eris~\cite{eris} and Clutch~\cite{clutch}, more tangentially to our work, \citet{QSL} developed a weakest precondition calculus for quantitative reasoning about probabilistic pointer programs in QSL, a quantitative analog of classical separation logic.
A different line of work develop separation logics in which separating conjunction models probabilistic independence.
This was first explored in probabilistic separation logic (PSL)~\cite{PSL} and subsequently extended to reason about conditional independence~\cite{bunched_logic_conditional_independence, lilac, bluebell} and negative dependence~\cite{SL_negative_dependence}.

\paragraph{Program Logics for Cryptographic Security}

CertiCrypt~\cite{pRHL, certicrypt} is a framework implemented in Coq~\cite{coq} for verifying code-based cryptographic proofs.
Programs in CertiCrypt are written in pWhile, a probabilistic imperative language, and the logic is based on pRHL~\cite{pRHL}.
CertiCrypt can prove approximate results such as the PRP/PRF lemma only at the level of the Coq meta-logic, since the program logic itself is based on exact couplings.

Building on pRHL and CertiCrypt, EasyCrypt~\cite{easycrypt} is a stand-alone tool for cryptography, integrating automation via SMT solvers. %
Although EasyCrypt can reason about simple rejection samplers~\cite{leakage-free-jasmin}, existing proofs require analysing the probability of each outcome, instead of a relational equivalence proof.
Rejection samplers with dynamic references or sophisticated early-abort, like the B+ tree sampler, would be difficult to do in this setting.

SSProve~\cite{SSProve, SSProve2} is a framework implemented in Coq~\cite{coq} for writing so-called state-separating proofs~\cite{ssp}.
Based on a monadic pRHL-like logic~\cite{RelPL}, games in SSProve are split into packages operating on disjoint states, which enables some amount of modular reasoning.
In contrast to the language considered by \theaplog, SSProve is first order and does not support dynamically allocated local state.

%% file: conclusion.tex
\section{Conclusion}
\label{sec:conclusion}

We presented \theaplog, the first higher-order separation logic for approximate
relational reasoning. In addition to establishing approximate bounds between
probabilistic programs, we developed a novel logical relation in \theaplog for
proving contextual refinement, by parameterizing over arbitrary positive error.
We demonstrated the strengths of \theaplog on various case studies involving
higher-order, local state, and non-trivial rejection sampling
behavior. Using \theaplog, we proved both approximate and exact examples, and also used the logical relation
to establish contextual refinements of examples that were previously out of scope.

 We believe \theaplog opens up numerous avenues for future work related to
 security of cryptographic protocols. Firstly, we would like to extend \theaplog
 to reason about concurrent programs in order to prove
 security guarantees of distributed systems. Secondly, it would be interesting to
 explore how \theaplog can be further improved to reason about time complexity of
 programs, to bound the computational power of adversaries.
 Finally, we aim to modify \theaplog to reason about several other security
 properties, including differential privacy and probabilistic sensitivity.

%% file: appendix-case-studies.tex
\section{Extended Case Studies}
\label{app:extended-case-studies}

\subsection{Simulating Dice}
\label{app:vnd}

Rejection samplers are a kind of Las Vegas algorithms used to simulate complicated probability distributions with simple probabilistic primitives.
These algorithms loop repeatedly, only terminating when it produces an acceptable value, one that correspond to a value in the target distribution.
Previously in \cref{sec:rej-samp}, we showed how to prove a simple rejection sampler is equivalent to a $\Rand$ expression.

To show that our techniques scale, we show a slightly more complicated implementation of a rejection sampler that simulates a 6-faced die roll (we assume faces are numbered from 0 to 5) with fair coin flips.
The na\"ive implementation would flip three coins, interpret the result as a binary number between 0 and 7, return the result if it is from 0 to 5, and restart the simulation otherwise.
A more efficient implementation would use an \emph{early abort} strategy: after observing the first two coin flips, if they are both 1 we can restart without the need for a third coin flip.
We implement an early abort rejection sampler $\dsim$ in Figure~\ref{fig:dice-algos}.
We will show that this is contextually equivalent to a uniform die roll, i.e., program $\droll$ in Figure~\ref{fig:dice-algos}.
As an intermediate step, we prove that they are both contextually equivalent to the simple rejection sampler $\drej$ in the same figure, which samples a uniform number between 0 and 7, returns the result if it is 5 or below, and restarts otherwise. 

\begin{figure*}[ht]
  \begin{minipage}{0.3\textwidth}
        \begin{align*}
	  &\dsim \ () \eqdef{}\\
          &\hspace{1em} \ghostcode{\Let \iota = \AllocTape 1 in }\\
          &\hspace{1em} \Let b2 =\Rand 1\ \ghostcode{\iota} in\\
          &\hspace{1em} \Let b1 =\Rand 1\ \ghostcode{\iota} in\\
	  &\hspace{1em} \If (b1 == 1~\&\&~b2 == 1) then \dsim\\
          &\hspace{1em} \Else \Let b0 =\Rand 1\ \ghostcode{\iota} in\\
          &\hspace{1em} \phantom{\Else} 4*b2 + 2*b1 + b0
    	\end{align*}
  \end{minipage}
  \begin{minipage}{0.3\textwidth}
        \begin{align*}
	  &\drej \ () \eqdef{}\\
          &\hspace{1em} \ghostcode{\Let \iota = \AllocTape 7 in }\\
          &\hspace{1em} \Let r =\Rand 7\ \ghostcode{\iota} in\\
	  &\hspace{1em} \If (r > 5) then \drej\\
          &\hspace{1em} \Else r
    	\end{align*}
  \end{minipage}
  \begin{minipage}{0.3\textwidth}
        \begin{align*}
	  &\droll \ () \eqdef{}\\
          &\hspace{1em} \ghostcode{\Let \iota = \AllocTape 5 in }\\
          &\hspace{1em} \Rand 5\ \ghostcode{\iota}
    	\end{align*}
  \end{minipage}
  \caption{Three algorithms to sample a fair die}
  \label{fig:dice-algos}
\end{figure*}

The proof thus requires showing the chains of logical refinements $\emptyset \proves \dsim \precsim \drej \precsim \droll \colon \tunit \to \tnat$ and  $\emptyset \proves \droll \precsim \drej \precsim \dsim \colon \tunit \to \tnat$.
The proofs of $\dsim \precsim \drej$ and $\drej \precsim \dsim$ are mostly symmetric.
The proof relies on using \ruleref{wp-rec} and the \ruleref{wp-many-to-one} coupling rule, to ensure that the three bits we sample to the tape with bound $1$ are a binary encoding of the number sampled to the tape with bound $7$, and therefore both conditionals resolve to the same branch.
In the case they both take the branch with the recursive call, we can apply our inductive hypothesis, otherwise both programs will terminate immediately and return equal values.

The proofs of $\drej \precsim \droll$ and $\droll \precsim \drej$ is almost identical to the proofs presented for the rejection samplers (see \cref{sec:rej-samp}), except that we are not only proving that the two programs are equivalent when executed in isolation, but \emph{contextually equivalent} under all contexts using our logical relations.
The former of the two uses \ruleref{wp-rec}, and our novel rule for fragmented couplings.
Note that $\droll$ only consume a single sample on the tape.
The rule for fragmented couplings ensures that we will either sample to the tapes a value above $5$ on the left and \emph{nothing} on the right, or we will sample the same value, between $0$ and $5$, to the tapes on both sides.
In the first case, $\drej$ will consume the value on the tape and call itself recursively, which allows us to use our inductive hypothesis.
In the second case, both programs will consume values on their tapes, read the same number, and return equal values.

Finally, we show $\droll \precsim \drej$. This proof cannot be done by applying \ruleref{wp-rec} since the program on the left is not recursive.
We will instead use induction by error amplification, through the rule \rref{log-ind-err}.
Let us set $k = 4/3$.
After applying this rule, we will get ownership of $\upto\err$ for some arbitrary $\err$, plus the induction hypothesis
\[ \upto{(4/3)\cdot\err} \wand (\emptyset \proves \droll~() \precsim \drej~() \colon \tnat) \,,\]
while our goal becomes
\[ \emptyset \proves \droll~() \precsim \drej~() \colon \tnat \,.\]
We now use our rule for fragmented couplings with errors.
This ensures that either (1) we sample to the tapes a value above $5$ on the right hand side, nothing on the left, and we amplify our credits by $4/3$, or (2) we sample identical values, between $0$ and $5$, to the tapes on both sides.
In the first case, the program on the right hand side will call itself recursively, but now we will own $\err{(4/3)\cdot\err}$, which is precisely what we need to apply our inductive hypothesis and conclude.
Otherwise, both values will have the same value on the tapes, and will terminate and return the same result.

While this example is conceptually simple, the reasoning patterns it uses, as well as the different induction principles that we can use depending on the presence or absence of recursion are important subtleties of our approach, and is fundamental to understanding other examples, such as that of the B+ tree in \cref{app:sampling-b-trees}.

\subsection{Sampling from B+ Trees}
\label{app:sampling-b-trees}

In this case study, we show the correctness of a rejection sampling scheme developed by \citet{b+_tree_sample} for drawing a random sample from a B+ tree.
Up to this point, previous examples have made use of only simple forms of state and the contextual equivalences were for simple type signatures.
This case study demonstrates how \theaplog is able to handle complex mutable state and establish equivalences that rely on type abstraction.

To motivate \citeauthor{b+_tree_sample}'s sampling algorithm, we first summarize some relevant facts about B+ trees.
A B+ tree~\citep{b+_tree} is a tree data structure that is widely used for storing data in filesystems and databases.
In contrast to a binary search tree, a B+ tree's internal nodes may have more than 2 children.
Random sampling from a B+ tree can be used to draw random records from such databases in order to carry out a statistical analysis.
Because the tree may store many elements, it is not efficient when drawing a sample to first reprocess the entire database into an alternate representation.
Instead, the sample must be drawn working directly over the tree structure.

The sampling algorithm we consider relies on 3 key properties of a B+ tree: (1) data elements are only stored at the leaves of the tree, (2) the height of the tree is perfectly balanced, meaning that the length of the path from the root to a leaf is the same for all leaves, (3) each node has at most $\maxchildnum$ children.
Since the algorithm only requires these properties for correctness, our proof will work with trees that are only assumed to satisfy these three properties, instead of assuming all of the invariants of a B+ tree.

Before presenting \citeauthor{b+_tree_sample}'s algorithm, let us first consider a na\"ive sampling algorithm that will serve as a correctness specification.
If we knew that the tree contained $N$ total elements, then one approach to drawing a random sample would be to first generate a random number uniformly from $\{0, \dots, N-1\}$ and then find and return the $i$-th element in the tree, numbering the leaves from left to right.
This approach correctly produces a uniform sample from the tree, but the challenge lies in efficiently finding the $i$-th element in the tree.
It is easy to find this element if we assume that the tree is a \emph{ranked} B+ tree, where intermediate notes are additionally annotated with the total number of leaves that are descendants of the node.
The function $\naiveprog$ in \cref{fig:naive-algo} implements this algorithm for sampling from a ranked B+ tree.
However, maintaining this rank information has a cost: \emph{every} insertion in the tree requires modifying all of the nodes that are ancestors of the inserted node to increase the recorded leaf counts.
In contrast, inserting into a (non-ranked) B+ tree, most insertions only require modifying the parent of the inserted element.

\begin{figure*}[t]
  \begin{minipage}[t]{0.5\linewidth}
        \newcommand{\gettotal}{\langv{num\_leaves}}
        \begin{align*}
          &\naiveprog\ \tree \eqdef{}\\
          &\hspace{1em} \ghostcode{\Let \iota = \AllocTape~(\gettotal\ \tree - 1) in }\\
          &\hspace{1em} \Let i=\Rand~(\gettotal\ \tree - 1)\ \ghostcode{\iota} in\\
          &\hspace{1em} \DumbLet \Rec f\ t\ \logv{num}= \\
          &\hspace{2em} \MatchML t with
          | \langv{Lf}\ v=> v 
          | \langv{Br}\ l=> {
            \begin{array}[t]{l}
              \\
            \mkern-46mu\Let (\logv{prev},\logv{idx}) = \langv{search}\ l\ \logv{num} in\\
            \mkern-46mu  \Let \logv{child} = l[\logv{idx}] in\\
            \mkern-46mu  f\ (\deref\logv{child})\ (\logv{num}-\logv{prev})
              \end{array}} 
            end{\In} \\
          &\hspace{1em} f\ \tree\ i
    \end{align*}
  \end{minipage}%
  \begin{minipage}[t]{0.44\linewidth}
        \begin{align*}
          &\optimizedprog\ \tree\eqdef{}\\
          &\hspace{1em} \DumbLet \Rec{} \optimizedsearch\ t\ \ghostcode{\iota} = \\
          &\hspace{2em} \MatchML t with
          | \langv{Lf}\ v=> \Some\ v 
          | \langv{Br}\ l=> {
            \begin{array}[t]{l}
              \Let \logv{idx}=\Rand~(\maxchildnum-1)~\ghostcode{\iota} in\\
              \MatchML {\listnth\ l\ \logv{idx}} with 
              | \Some \logv{child} => \optimizedsearch\ (\deref\logv{child})
              | \None => \None
              end{}
              \end{array}}
          end{\In}\\
          &\hspace{1em} \DumbLet \Rec{} f~\_= \\
          &\hspace{2em} \ghostcode{\Let \iota = \AllocTape~(\maxchildnum-1) in}\\
          &\hspace{2em} \MatchML \optimizedsearch\ \tree\ \ghostcode{\iota} with
          | \Some\ v=> v 
          | \None => f\ \TT
            end{\In}\\
          &\hspace{1em}f~\TT
    \end{align*}
  \end{minipage}
  \caption{Na\"ive algorithm for sampling from a ranked B+ tree and the \citet{b+_tree_sample} algorithm for rejection sampling from a non-ranked B+ tree.}
  \label{fig:naive-algo}
\end{figure*}

\citeauthor{b+_tree_sample} developed a rejection sampling algorithm for sampling from a non-ranked B+ tree.
We call this the optimized algorithm, implemented as $\optimizedprog$ in \cref{fig:naive-algo}.
This function makes use of the early abort technique we saw in \cref{app:vnd}.
Starting from the root, it samples a random number $i$ uniformly from $\{0, 1, \dots, \maxchildnum-1\}$, where $\maxchildnum$ is the maximum number of children a node can have.
It selects the $i$-th child and recurses on it, until it reaches a leaf.
If the current node does not have an $i$-th child, we return to the root and restart the algorithm.
The intuition behind the correctness of the optimized algorithm is that it is somewhat similar to sampling random leafs from a full multi-way tree, i.e.\ a B+ tree where each intermediate node holds $\maxchildnum$ branches.
In the case where we walk down a branch that is not present in the original B+ tree, we reject this branch and start all over again.

Our main result for this case study shows that the na\"ive algorithm and the optimized algorithm are equivalent.
Of course, these algorithms are only equivalent when they operate over well-formed trees, so we state this result as a contextual equivalence about two different implementations of an abstract data type with operations for constructing and sampling from the tree.
To state this precisely, we first define the following functions (code omitted): $\inittree$, which takes an integer and returns a B+ tree containing that single integer, $\inserttree$, which inserts an integer into a tree, and $\buildranked$, which takes a (non-ranked) B+ tree and returns a ranked B+ tree with the same entries and shape.
Next, we define the following two packed tuples that bundle the operations for the B+ tree:
\begin{align*}
\optpack &\eqdef \Pack (\inittree, \inserttree, \optimizedprog) \\
\naivepack &\eqdef \Pack (\inittree, \inserttree, \lambda t. \naiveprog\ (\buildranked\ t))
\end{align*}
where the sampling routine in $\naivepack$ takes a tree $t$, builds the ranked version of the tree, and then uses the na\"ive routine.\footnote{Of course it would be highly inefficient to construct a ranked tree every time a sample is to be drawn, but $\naivepack$ here serves as a correctness specification for $\optpack$, so its efficiency is not relevant.}
With these preliminaries in place, our main result can be stated as $\ctxeq{}{\optpack}{\naivepack}{\treetype}{}$, where $\treetype$ is the following existential type:
\[ \treetype \eqdef \Exists \tau. (\intty \rightarrow \tau) \times (\tau \times \intty \rightarrow \unitty) \times (\tau \rightarrow \intty) \]
The proof of this equivalence in our full Coq development is too long to fully explain here.
However, at a high level, the proof has two components.
First, there is the non-probabilistic reasoning showing that the various routines traverse and modify the trees correctly, \eg that the height-balanced invariant is maintained by $\inserttree$, or that $\buildranked$ correctly computes ranks.
This aspect in fact makes up the bulk of the proof, and consists of using traditional separation-logic style reasoning about trees.
For this part, \theaplog{}'s support for the rich reasoning principles developed in earlier separation logics is essential.

The second component is the actual probabilistic reasoning using couplings.
Here, the coupling reasoning in this proof is quite similar to the arguments used for proving the equivalence of the die sampling routines in \cref{app:vnd}.
To make this correspondence clearer, and to simplify the reasoning, we introduce an intermediate sampling routine, $\intermediateprog$, shown in \cref{fig:intermediate-algo}.
The intermediate program takes in a tree, computes the depth $\depth$ of its leaves, and samples a value from $\Rand(\maxchildnum^{\depth}-1)$.
It then interprets this number as a path through the tree.
To do so, the program treats it after it were a $\depth$ digit number written in base $\maxchildnum$, in which the $i$-th digit represents a child to select at depth $i$. 
If, on reaching depth $i$ it finds that the corresponding child does not exist, then it rejects and repeats with a fresh sample.

\begin{figure*}[t]
  \centering
      {\begin{align*}
          &\intermediateprog\ \tree \eqdef{}\\
          &\hspace{1em} \Let \depth = \getdepth\ \tree in \\
          &\hspace{1em} \Let \lbl = \AllocTape~(\maxchildnum \expsymb{} \depth -1) in \\
          &\hspace{1em} \DumbLet \Rec{} \intermediateprog'\ t= \\
          &\hspace{2em} \Let \logv{idx} = \Rand~(\maxchildnum\expsymb{}  \depth -1)~\lbl in \\
          &\hspace{2em} \DumbLet \Rec{} f\ t\ \logv{num}\ d= \\
          &\hspace{3em} \MatchML t with
          | \langv{Lf}\ v=> v 
          | \langv{Br}\ l=> {
            \begin{array}[t]{l}
              \Let \logv{idx}=\logv{num}\ \text{\`{}quot\`{}}\ (\maxchildnum\expsymb{}(d-1)) in\\
              \MatchML {\listnth\ l\ \logv{idx}} with 
              | \Some \logv{child} => f\ (\deref\logv{child})\ (\logv{num}-\logv{idx}\cdot(\maxchildnum\expsymb{} (d-1)))\ (d-1)
              | \None => \intermediateprog'\ \tree
              end{}
              \end{array}}
            end{\In} \\
          &\hspace{3em}{f~t~\logv{idx}~\depth \In} \\
          &\hspace{1em}\intermediateprog'~\tree
    \end{align*}
  }
  \caption{Intermediate algorithm for sampling from a non-ranked B+ tree.}
  \label{fig:intermediate-algo}
\end{figure*}

We can see then that $\naiveprog$ is similar to $\droll$: it always succeeds because it samples an index of a valid leaf, just as $\droll$ always samples a number that is in range.
Meanwhile, $\intermediateprog$ is like $\drej$, as it samples a large number representing an entire path in a tree, and then rejects if that path is invalid, much as $\drej$ rejects if its sample is too large.
Finally, $\optimizedprog$ is like $\dsim$, in that it samples the path layer-by-layer and rejects early if the path is invalid, just as $\dsim$ samples bit-by-bit and rejects early if the number is already too large.
Thus for example, in proving that $\naiveprog \precsim \intermediateprog$, we use fragmented couplings and error amplification, just as we did for $\droll \precsim \drej$, while proving $\optimizedprog \precsim \intermediateprog$, we use the \ruleref{wp-many-to-one} and \ruleref{wp-rec} rule.